\documentclass[trackchanges, numerical]{aastex701}

\usepackage{amsmath,amsfonts,amssymb}
\usepackage{graphicx}
\usepackage{tocloft}
\usepackage[T1]{fontenc}
\usepackage{siunitx} 
\usepackage{subcaption}
\usepackage{booktabs}
\usepackage{multirow}
\usepackage{array}    
\usepackage[shortlabels]{enumitem}
\usepackage{comment}
\usepackage{caption}
\usepackage{svg}
\usepackage{float}

\usepackage{booktabs}
\usepackage{tabularx}
\usepackage{siunitx} 

\begin{document}

\citestyle{numeric}

\title{Prototyping of 6.2-mm-Pitch Fiber Positioner Modules for Stage-V Telescope Instrumentation}

\author[0000-0002-2323-8332]{Malak Galal}
\affiliation{École Polytechnique Fédérale de Lausanne, Laboratory of Astrophysics (LASTRO), Lausanne, Switzerland}
\email{malak.galal@epfl.ch}

\author[0009-0004-1243-2545]{Maxime Rombach}
\affiliation{École Polytechnique Fédérale de Lausanne, Laboratory of Astrophysics (LASTRO), Lausanne, Switzerland}
\email{maxime.rombach@epfl.ch}

\author[0009-0007-8997-896X]{Jonathan Wei}
\affiliation{École Polytechnique Fédérale de Lausanne, Laboratory of Astrophysics (LASTRO), Lausanne, Switzerland}
\email{}

\author[0009-0001-4773-3354]{Oliver Pineda Suárez}
\affiliation{École Polytechnique Fédérale de Lausanne, Laboratory of Astrophysics (LASTRO), Lausanne, Switzerland}
\email{}

\author[]{Ricardo Araújo}
\affiliation{École Polytechnique Fédérale de Lausanne, Laboratory of Astrophysics (LASTRO), Lausanne, Switzerland}
\email{}

\author[]{Sébastien Pernecker}
\affiliation{École Polytechnique Fédérale de Lausanne, Laboratory of Astrophysics (LASTRO), Lausanne, Switzerland}
\email{}

\author[0000-0002-9964-1005]{Abby Bault}
\affiliation{Lawrence Berkeley National Laboratory (LBNL), Berkeley, California, United States of America}
\email{}

\author[0000-0002-3461-0320]{Joseph Harry Silber}
\affiliation{Lawrence Berkeley National Laboratory (LBNL), Berkeley, California, United States of America}
\email{}

\author[0009-0007-6668-6964]{Nicholas Wenner}
\affiliation{Lawrence Berkeley National Laboratory (LBNL), Berkeley, California, United States of America}
\email{}

\author[]{Robert Besuner}
\affiliation{Lawrence Berkeley National Laboratory (LBNL), Berkeley, California, United States of America}
\email{}

\author[0000-0002-8828-5463]{David Kirkby}
\affiliation{Department of Physics and Astronomy, University of California, Irvine, United States of America}
\email{}

\author[]{William Van Shourt}
\affiliation{Department of Physics and Astronomy, University of California, Irvine, United States of America}
\email{}

\author[]{Stefane Caseiro}
\affiliation{Micro Precision Systems AG (MPS), Bienne, Switzerland}
\email{}

\author[]{Corentin Magnenat}
\affiliation{Micro Precision Systems AG (MPS), Bienne, Switzerland}
\email{}

\author[]{Yves Moser}
\affiliation{Micro Precision Systems AG (MPS), Bienne, Switzerland}
\email{}

\author[]{Yasuyuki Kobayashi}
\affiliation{Orbray Ltd, Aomori, Japan}
\email{}

\author[]{Eri Fukushima}
\affiliation{Orbray Ltd, Aomori, Japan}
\email{}

\author[]{Satoshi Sonoda}
\affiliation{Orbray Ltd, Aomori, Japan}
\email{}

\author[]{Ayumu Suto}
\affiliation{Orbray Ltd, Aomori, Japan}
\email{}

\author[0000-0002-5042-5088]{David J. Schlegel}
\affiliation{Lawrence Berkeley National Laboratory (LBNL), Berkeley, California, United States of America}
\email{}

\author[0000-0002-4616-4989]{Jean-Paul Kneib}
\affiliation{École Polytechnique Fédérale de Lausanne, Laboratory of Astrophysics (LASTRO), Lausanne, Switzerland}
\email{}

\begin{abstract}

Small-pitch populated focal planes are essential enabling technologies for the next generation of highly multiplexed astronomical instruments. As modern astrophysics relies on massive spectroscopic surveys to study dark energy, dark matter, and galactic assembly, the ability to observe thousands of targets simultaneously has become paramount. To achieve these ambitious scientific goals, optical fibers must be packed into the telescope's focal plane with unprecedented density and accuracy.

This work reports on comprehensive prototyping activities for novel 6.2 mm-pitch alpha–beta (theta–phi) fiber positioner modules. Achieving reliable operation at this extremely miniaturized scale presents formidable mechanical and control-system challenges. We provide a detailed comparative analysis of two primary architectural approaches: trillium-based mechanisms and independently actuated robotic designs. A rigorous quantitative assessment was conducted for both prototype models. Critical metrics such as XY positioning repeatability, non-linearity, and gear backlash were evaluated, as these directly dictate the targeting accuracy of the fiber on the sky. Furthermore, we analyzed fiber tilt angles, a crucial factor given its severe implications for Focal Ratio Degradation and the subsequent loss of optical throughput to the spectrographs. Our analysis contextualizes these mechanical constraints with their direct implications for overall instrument performance and survey efficiency. Initial results are highly encouraging, indicating that these miniaturized positioners can successfully overcome spatial limitations while maintaining stringent tolerances. These promising metrics demonstrate that 6.2 mm-pitch modules are highly suitable for the next generation telescopes and the massive multi-object spectroscopic facilities.

\end{abstract}


\section{Introduction}
\label{sect:intro}  
The 3-D mapping of the Universe with the Sloan Digital Sky Survey (SDSS) (\cite{york_sloan_2000}), a 20-year endeavor of massive spectroscopy using the 2.5-m Sloan Telescope, has spurred new cosmological spectroscopic surveys for detailed 3-D mapping of the Universe. The large scale structure distribution of galaxies is a key measurement in mapping the matter in the Universe and probing its expansion, its content and puts constraints on the cosmological model and fundamental physics (\cite{percival_measuring_2007}).

Already started in 2020, the Dark Energy Spectroscopic Instrument (DESI) project (\cite{levi_desi_2013, collaboration_desi_2016}) has been surveying the Universe over the past four years. The first year of data indicates a possible deviation from the standard Lambda cold dark matter (LCDM) model (\cite{collaboration_desi_2024}). DESI will release at the end of 2025 data with about 35-million redshifts of galaxies (at z<1.6) and quasars, further refining current measurements. The ESA Euclid mission (launched in July 2023) (\cite{laureijs_euclid_2024}) and the ESO 4MOST project (first light in September 2025) (\cite{jong_4most_2012}) will complement the DESI survey by collecting together an additional 50+8 million redshifts out to z$\sim$2 (Euclid) and over the southern hemisphere (4MOST), respectively.

Although we do expect a significant improvement in cosmological measurement from these 3 major experiments, the 3-D mapping of the Universe will still be incomplete, as the redshift range 2<z<5 will be only sparsely mapped by those surveys.
\begin{figure}[!t]
     \centering
     \captionsetup[subfigure]{justification=centering}
     \captionsetup{justification=centering}
     \begin{subfigure}[b]{0.59\textwidth}
         \centering
         \includegraphics[width=\textwidth]{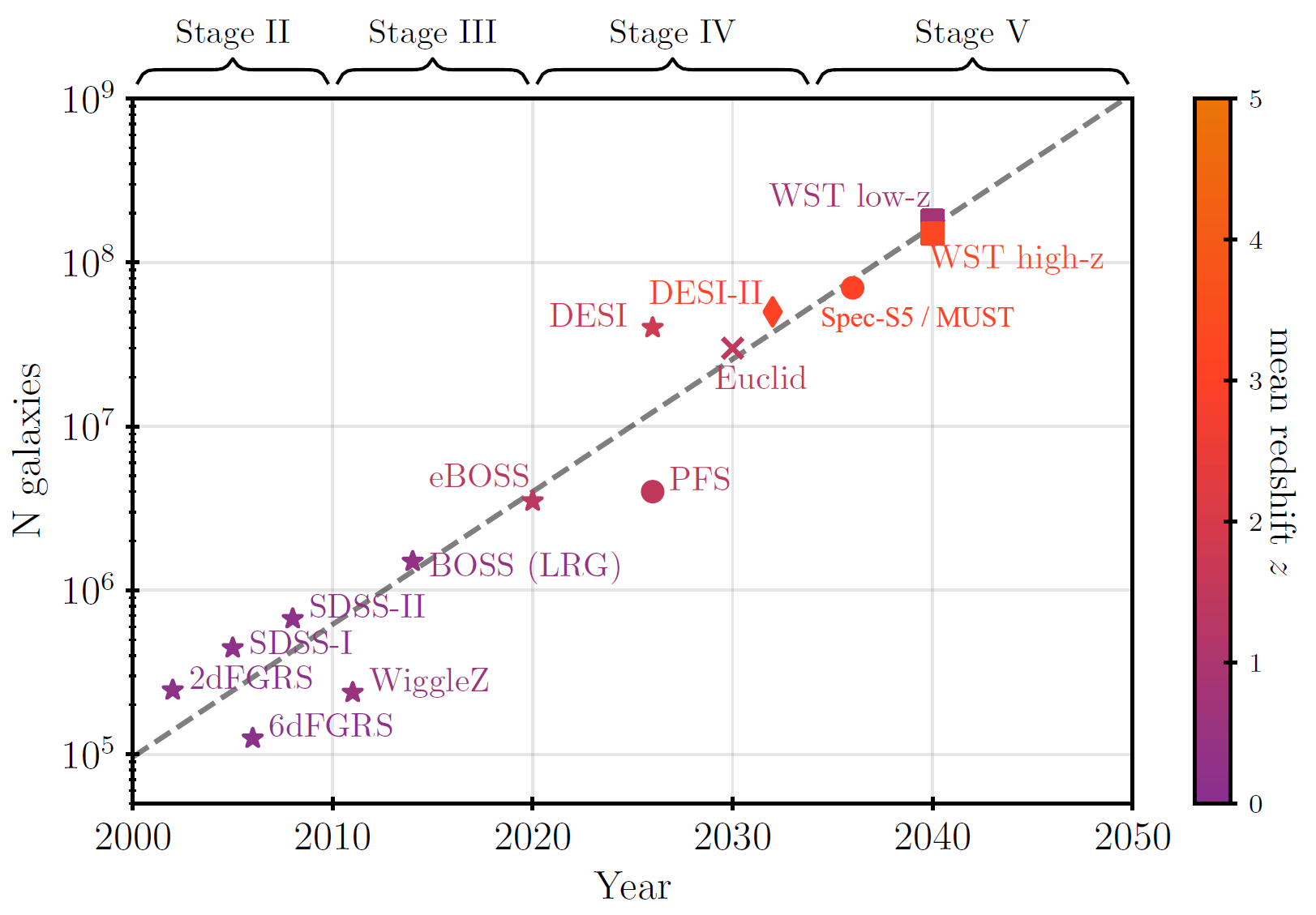}
         \caption{}
         \label{fig:time_vs_galaxies_surveyed}
     \end{subfigure}
     \hfill
     \begin{subfigure}[b]{0.39\textwidth}
         \centering
         \includegraphics[width=\textwidth]{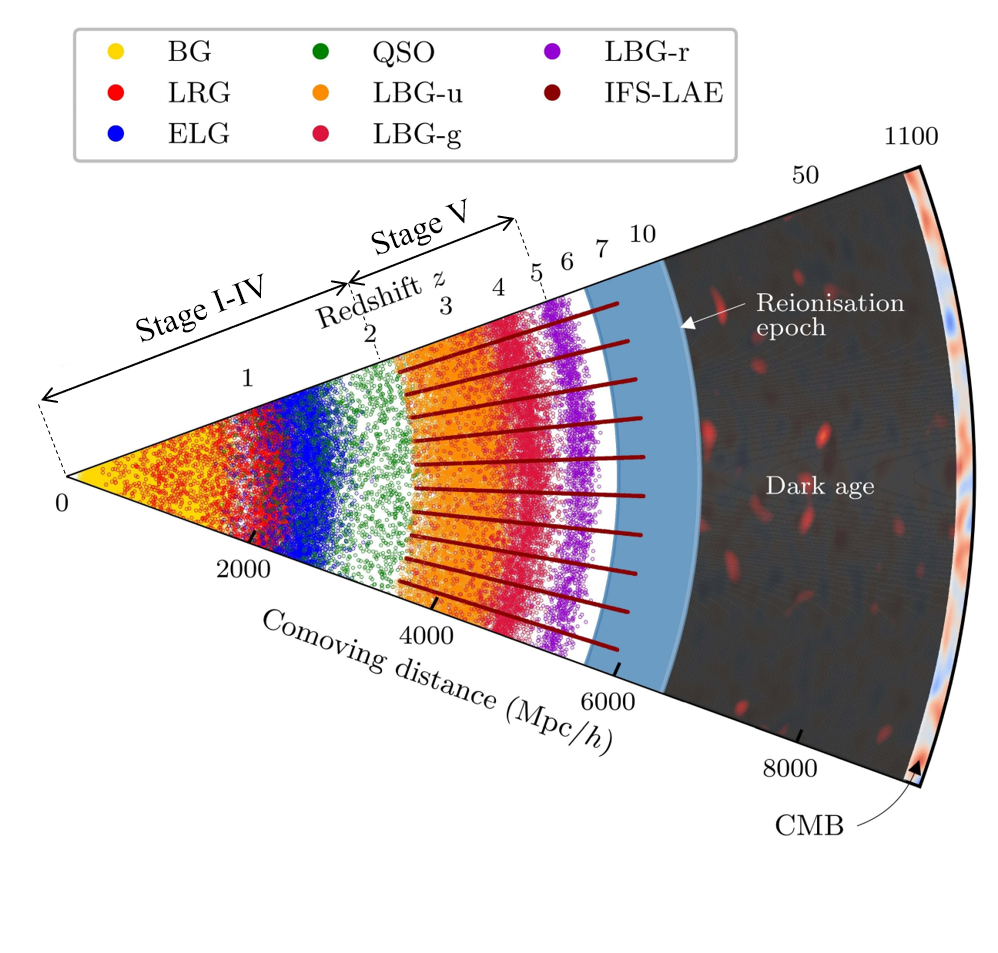}
         \caption{}
         \label{fig:redshifts_schematics}
     \end{subfigure}
        \caption{Graphs highlighting the interest of the astronomical community to pursue building Stage-V instruments to enhance our comprehension of the Universe; (a) Progress in wide cosmology spectroscopic redshift surveys of the year 2000 to 2050 (\cite{mainieri_wide-field_2024}); (b) Schematic using the light cone representation to show the range of future Stage-V MOS instruments (\cite{mainieri_wide-field_2024}).}
        \label{fig:stageV_surveys}
\end{figure}
This high redshift domain is, however, a key epoch for the formation of galaxies and has enough cosmological volume to detect the imprint of Inflation (the exponential expansion epoch of the Universe 10$^{-34}$ seconds after the Big-Bang) in the galaxy distribution, as well as to decisively constrain the neutrino masses, and further inform us on the nature of Dark Matter and Dark Energy. A number of international projects are getting ready to tackle this missing 3-D mapping of the Universe: Spec-S5 (led by Lawrence Berkeley National Laboratory) (\cite{schlegel2019astro2020, blanc_megamapper_2022, besuner_spectroscopic_2025}), MUST (led by Tsinghua University) (\cite{zhang_conceptual_2023, zhao_multiplexed_2025}), and the European Southern Observatory (ESO)-led Wide-field Spectroscopic Telescope (WST) (\cite{lee_wst_2024, mainieri_wide-field_2024}).

One of the main contributors to the success of the SDSS-V and DESI surveys is the implementation of the Optical Fiber Positioning Robots (\cite{pogge_robotic_2020, sanchez-gallego_multi-object_2020, poppett_overview_2024}). They lie at the heart of the telescope (the focal plane) and can be considered to be the nervous system of the whole structure. The light collected from the sky by the telescope's optical train is transmitted all the way through the optical fibers to reach the spectrographs. Since optical fibers need to be placed and actuated precisely to point at the desired targets, robotic positioners have been proposed and have proven to be the ultimate solution. SDSS-V relied on a total of 1,000 alpha-beta\footnote{We refer generally to fiber robots with parallel rotating axes as "alpha-beta" here. Some projects (e.g. DESI) use the nomenclature "theta-phi". In either case, the kinematics consist of a central axis (the angle "alpha" or "theta") upon which an eccentric axis (the angle "beta" or "phi") is mounted. This provides the necessary 2 degrees of freedom to cover a planar patrol zone.} robotic positioners (500 on each hemi-sphere) to complete its survey. DESI (\cite{silber_robotic_2023}) is currently employing 5,000 alpha-beta robots at a miniaturized 10.4 mm mounting pitch, successfully achieving the largest 3-D map of the observable Universe thus far. 
The success of these surveys has ignited great interest in expanding the number of robotic fiber positioners, to observe even more objects simultaneously.

Science requirements of future projects are typically looking to expand the number of robotic positioners by approximately a factor of 4x-5x, from the current state of the art (1-5,000 individually positioned fibers) to roughly 20-25,000. The baseline of 20-25,000 fibers across the focal plane is dictated by the 
survey speed and cosmological requirements of next-generation Stage-V spectroscopic instruments. As detailed for the Spec-S5 project \cite{schlegel2019astro2020, besuner_spectroscopic_2025}, high multiplexing is essential to efficiently map massive galaxy volumes at high 
redshifts ($2 < z < 5$). This capacity provides the survey speed necessary to constrain fundamental parameters like primordial non-Gaussianity ($f_{\rm nl}$) within a realistic multi-year observing schedule.

For this massive number of positioners to be placed in roughly the same-sized focal surface as previous surveys to match with the dimensions of the rest of the telescope components (e.g. the optical train), the robotic positioners themselves need to be made as small as possible. In 2022, Silber et al. (\cite{silber_25000_2022}) described a system architecture based on triangular "raft" modules, each containing 75 fiber robots mounted at 6.2 mm spacing, and showed initial prototypes that demonstrated feasibility of this miniaturization. In 2024, Rombach et al. (\cite{rombach_investigations_2024}) provided a rigorous study of focal plane coverage optimization and assembly of such modules, leading to a reduced module size recommendation of 63 robots.

In this paper we describe in-depth prototyping efforts of miniaturized robots for this modular concept, done by a collaboration of the Astrobots team at EPFL and a group of staff engineers and scientists at LBNL / UC Berkeley. We procured prototypes from two manufacturers: the Swiss company Micro Precision Systems (MPS) and the Japanese company Orbray. Each manufacturer assembled two prototype modules, with each module having 6 robotic positioners. The robots were mounted at 6.2 mm pitch, and the modules were sized for 63 robots, but with only 6 positions filled. The two manufacturers based their prototypes on two different alpha-beta style designs, with significant iteration and input from the engineering teams at EPFL and LBNL. EPFL and LBNL procured one module each from the two manufacturers. In this paper, we discuss the different modular designs on which the prototypes have been based, and present first results of the performance of both prototypes. 

\begin{figure}[!t]
    \captionsetup{justification=centering, margin = 2cm}
    \centering
    \includegraphics[width=0.5\linewidth]{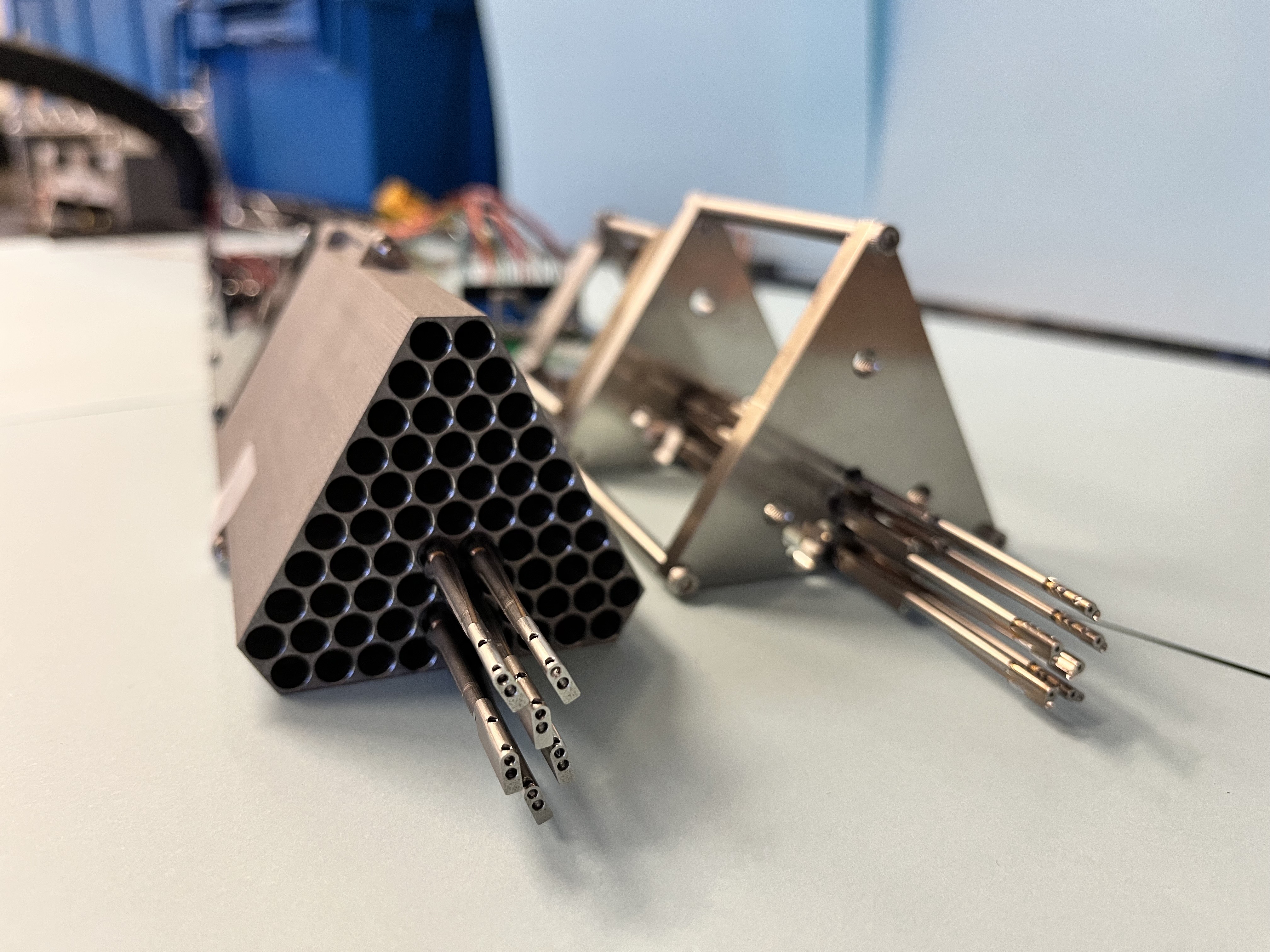}
    \caption{Two prototypes of modules from MPS (left) and Orbray (right) captured at LBNL for the project Spec-S5 (\cite{besuner_spectroscopic_2025})}
    \label{fig:MPS_orbray_modules}
\end{figure}
\section{The Modular Solution}
\label{sec:The Modular Solution}
\begin{figure}[!t]
     \centering
     \captionsetup[subfigure]{justification=centering}
     \begin{subfigure}[b]{0.52\textwidth}
         \centering
         \includegraphics[width=\textwidth]{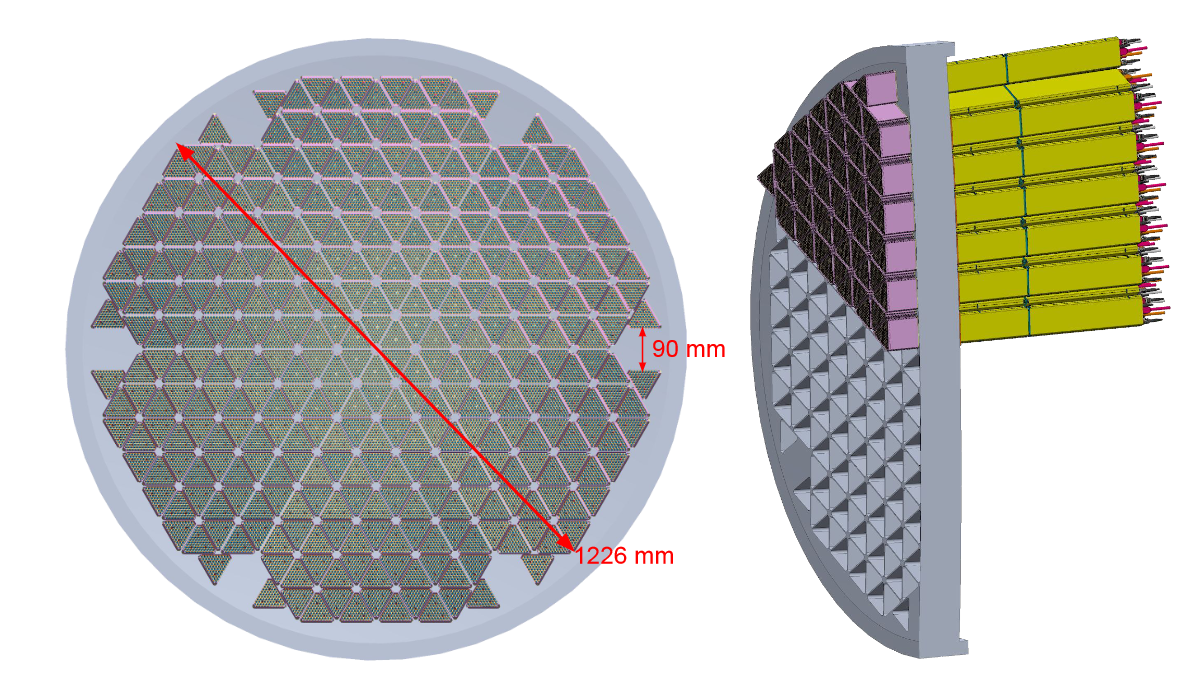}
         \caption{Modules integration in focal plate concept}
         \label{fig: module assembly in plate concept}
     \end{subfigure}
     \hfill
     \begin{subfigure}[b]{0.4\textwidth}
         \centering
         \includegraphics[width=\textwidth]{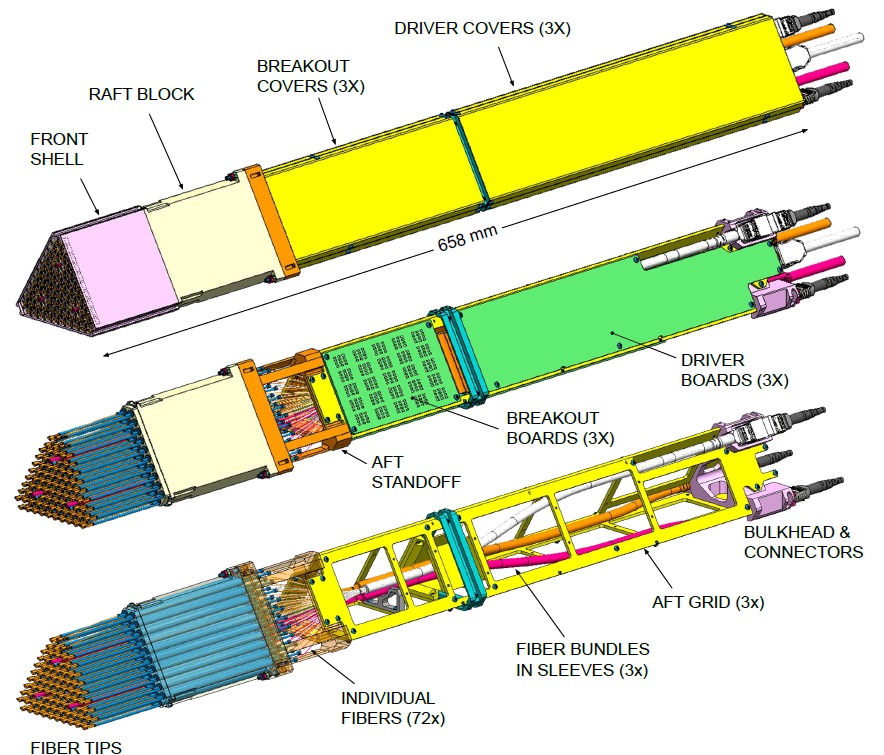}
         \caption{Module architecture}
         \label{fig:module concept architecture}
     \end{subfigure}
        \caption{Modular concept as described in Silber et al. (\cite{silber_25000_2022})}
        \label{fig:modular concept}
\end{figure}

Past experiences of multi-object spectrograph (MOS) instruments have shown the integration complexity of individual fiber positioner assembly, testing, and validation. MOONS (\cite{gonzalez_moons_2022}) and SDSS-V (\cite{pogge_robotic_2020}), with 1001 and 2 x 500 positioners, respectively, have chosen to individually assemble their robots into a single focal plate structure. Others have opted to segment the assembly of robots in their focal plane to parallelize unit testing, mutualize power supply, or fiber management. For instance, 4MOST fiber positioners are assembled on common linear structures of a number of robots that are individually tested prior to installation in their hexagonal layout (\cite{haynes_4most_2014}). DESI, on the other hand, segmented its circular robots layout in 10 individual \textit{petals} of 500 positioners (\cite{collaboration_desi_2024}). Those examples have shown the strength of modularizing focal planes for increasing numbers of fiber positioners. When scaling up to $\sim$20,000 for future Stage-V projects such as Spec-S5, MUST or WST, one of the key lessons learned from DESI was that module size should be selected not only for considerations of final assembly and operational maintenance, but also during development and manufacturing to "\textit{lower barriers to integrated tests}"
(\cite{silber_robotic_2023}). Therefore, the fiber positioners effort for future highly-multiplexed MOS instruments aims to construct standalone \textit{modules} (or \textit{rafts}) that will encapsulate a fixed number of positioners, easing pre-integration unit testing and on-site maintenance. It has been conceptualized in Silber et al. (\cite{silber_25000_2022}) as triangular units of 75 alpha-beta (or theta-phi) positioners, see Figure \ref{fig:modular concept}. Recent investigations (\cite{rombach_investigations_2024}) have shown that 63 positioners per module seemed more optimal for mechanical assembly, electronics distribution and focal surface matching. Details on the two module designs tested in this article can be found in Section \ref{sec:Different Designs for the Fiber Positioner Modules}.

\section{Different Designs for the Fiber Positioner Modules}
\label{sec:Different Designs for the Fiber Positioner Modules}

\begin{figure}[H]
    \centering
    \includegraphics[width=0.3\textwidth]{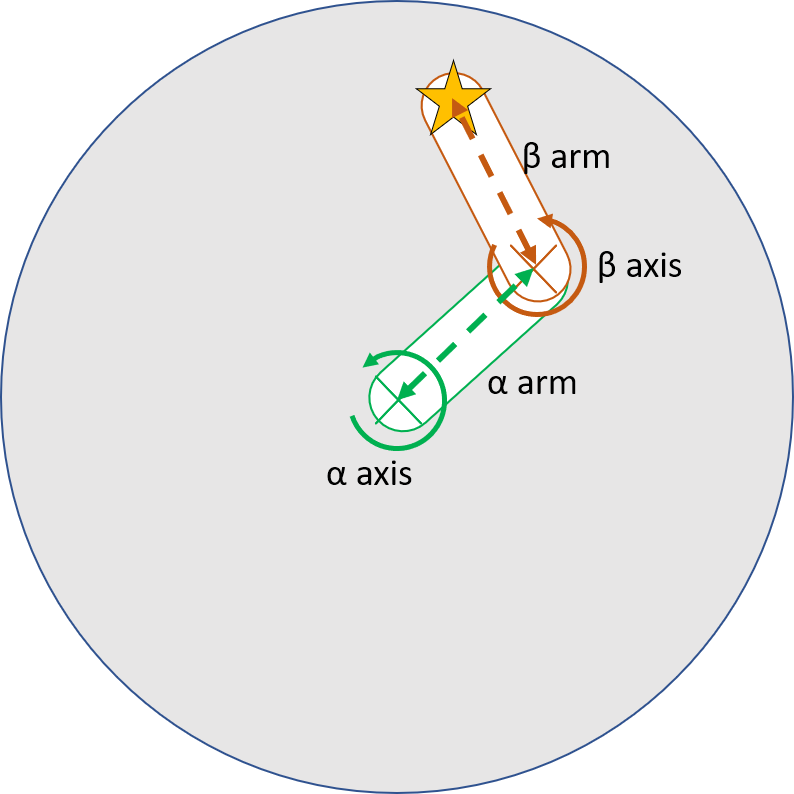}
    \caption{Schematic of a "alpha-beta" robot kinematics; the gray area is the patrol area into which the positioner can place its optical fiber - 63 of those patrol areas are combined in a module as shown in Figure \ref{fig:module_coverage}}
    \label{fig:robot_kinematics}
\end{figure}

\begin{figure}[!t]
     \centering
     \captionsetup[subfigure]{justification=centering}
     \begin{subfigure}[c]{0.31\textwidth}
         \centering
         \includegraphics[width=\textwidth]{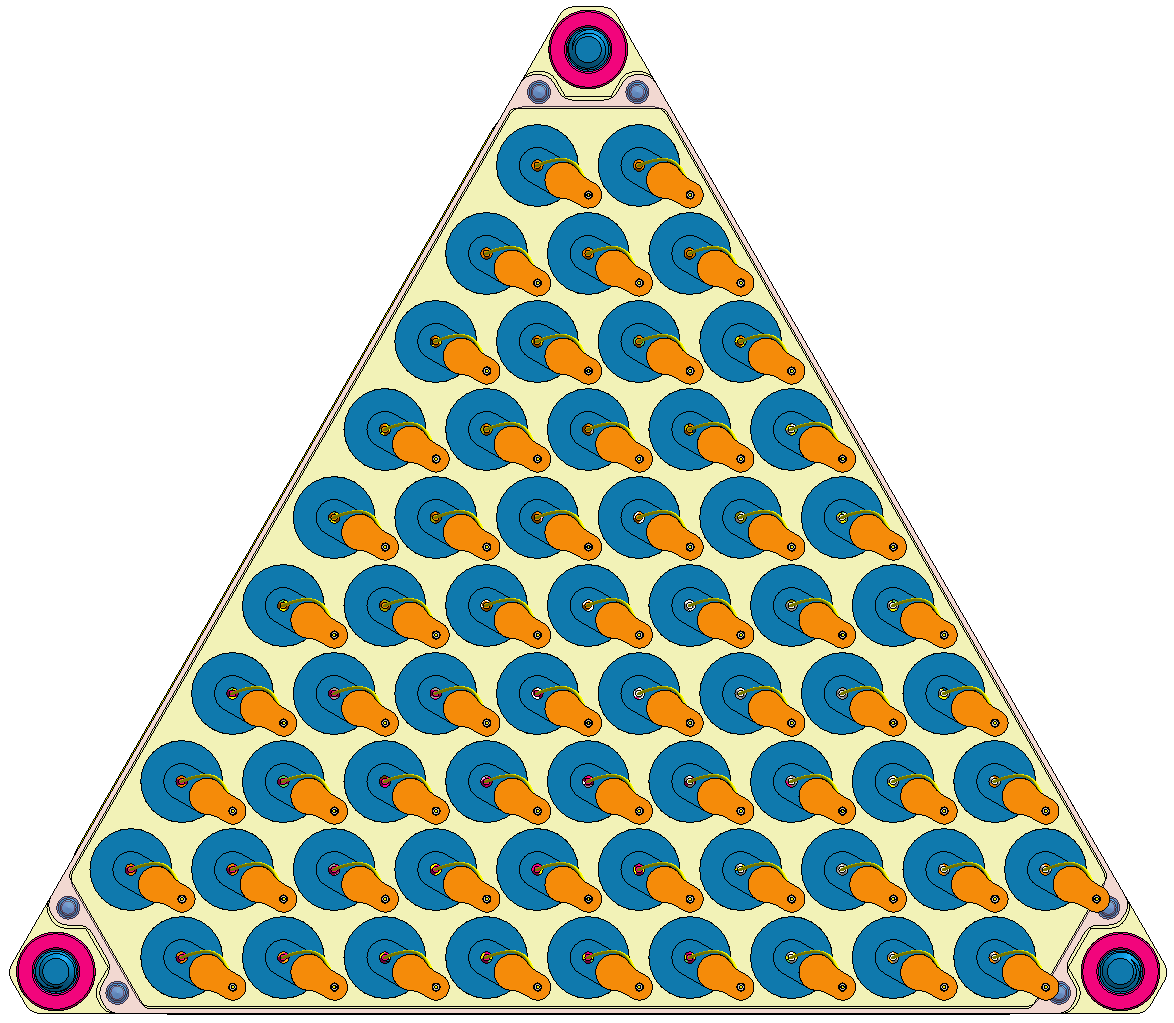}
         \caption{}
         \label{fig:indiv_robots}
     \end{subfigure}%
          \hfill
     \begin{subfigure}[c]{0.31\textwidth}
         \centering
         \includegraphics[width=\textwidth]{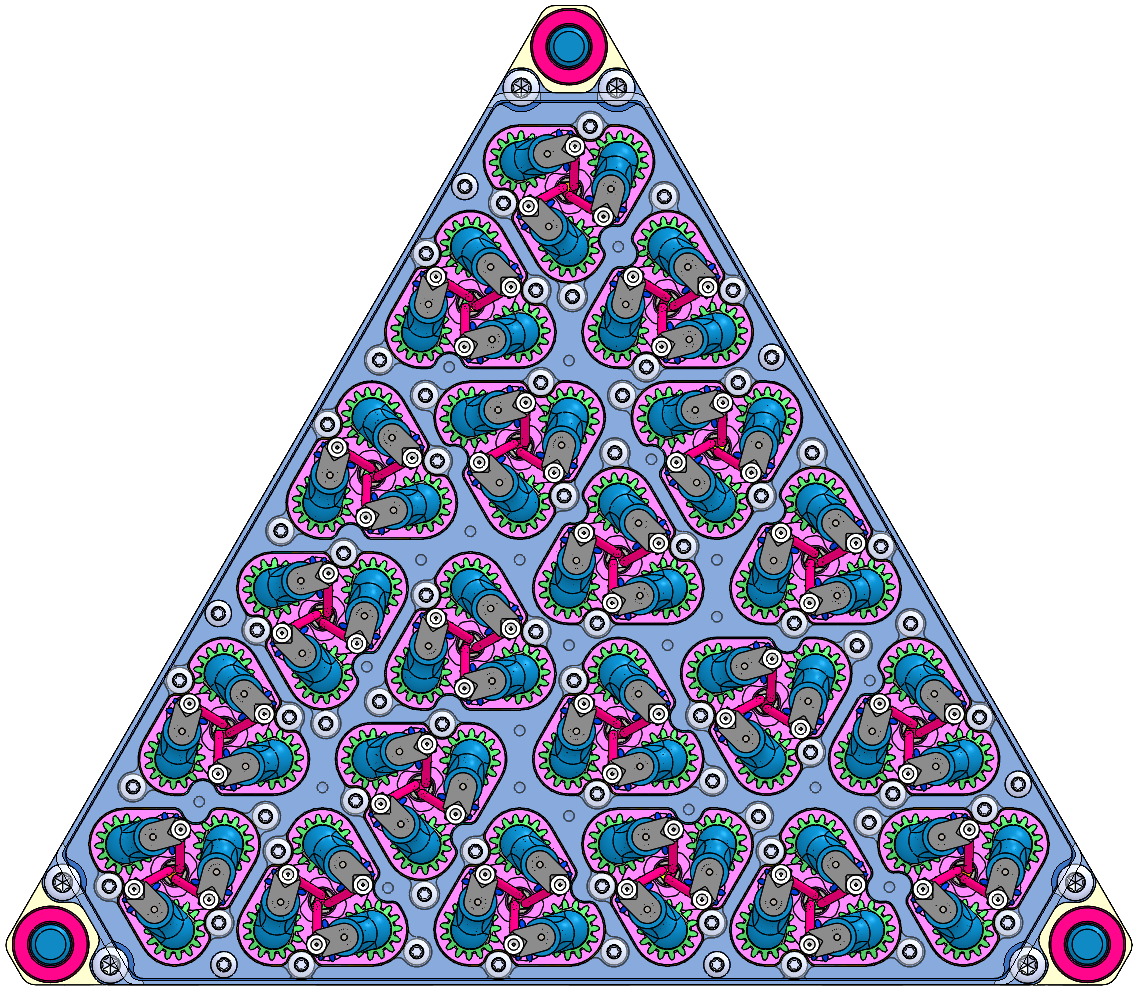}
         \caption{}
        \label{fig:trillium_robots}
     \end{subfigure}
      \begin{subfigure}[c]{0.31\textwidth}
         \centering
         \includegraphics[width=\textwidth]{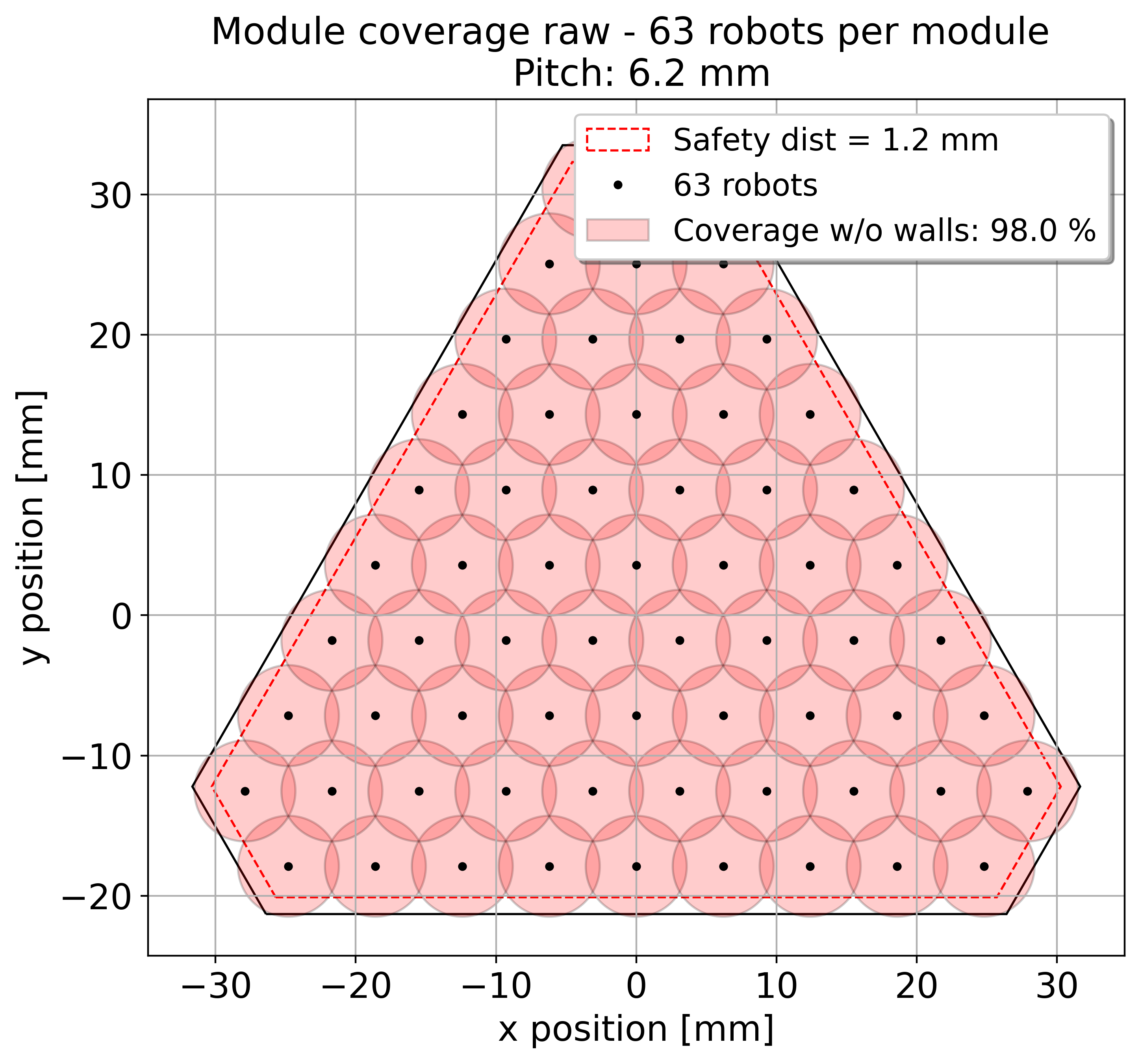}
         \caption{}
         \label{fig:module_coverage}
     \end{subfigure}%
        \caption{Different module architectures; (a) Individual robots architecture; (b) Trillium robots architecture; (c) Module coverage of both architectures.}

\end{figure}

The fiber positioner modules are intended to each host 63 robots with a pitch (i.e. center-to-center distance), of 6.2 mm. The robotic architecture is a SCARA (as seen in Figure \ref{fig:robot_kinematics}) similar to DESI, MOONS or SDSS-V with two axes of rotation: alpha-beta, or theta-phi. The robot's alpha and beta arms are the same length: 1.8 mm. The individual patrol area for each fiber is therefore a disk of 7.2 mm diameter. Within a given module, patrol areas overlap by up to 1.0 mm. (Figure \ref{fig:module_coverage}).

The two prototypes studied in this paper both have alpha-beta output kinematics, where each robot is driven by a pair of independent, DC brushless gear motors. The designs differ in the internal mechanics which convert gear motor shaft rotations to alpha and beta axis rotations of the fiber-holding arms.

The MPS raft implements an \textit{individual robots} design, in which (Figure \ref{fig:indiv_robots}) the robots are integrally built into the triangular chassis. The alpha and beta axes for each robot move independently from one another, i.e. the alpha arm can be rotated without moving its beta arm.

\begin{figure}[!b]
     \centering
     \captionsetup[subfigure]{justification=centering}
     \begin{subfigure}[c]{0.31\textwidth}
         \centering
         \includegraphics[width=\textwidth]{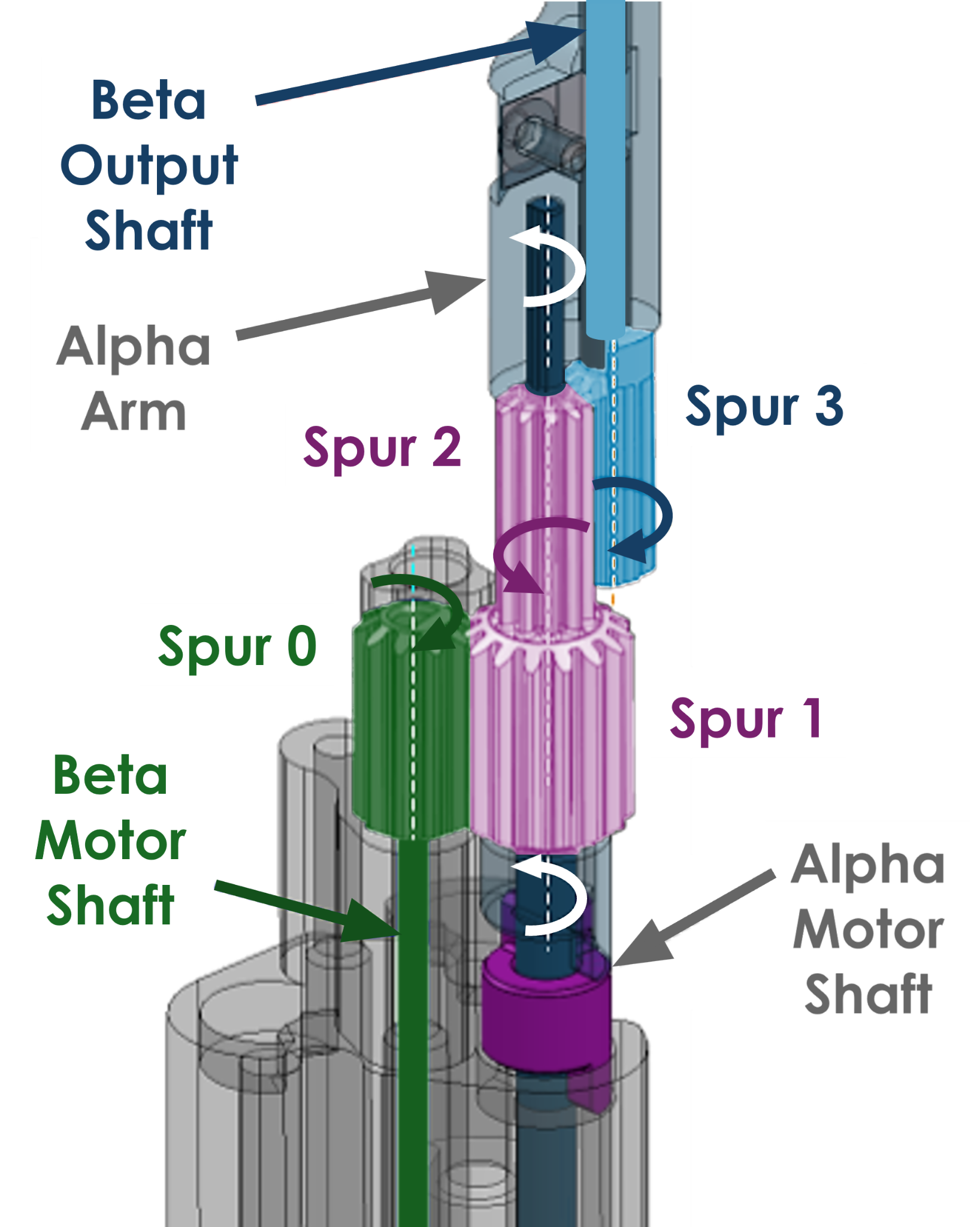}
         \caption{}
        \label{fig:trillium_mechanics}
     \end{subfigure}
      \begin{subfigure}[c]{0.31\textwidth}
         \centering
         \includegraphics[width=\textwidth]{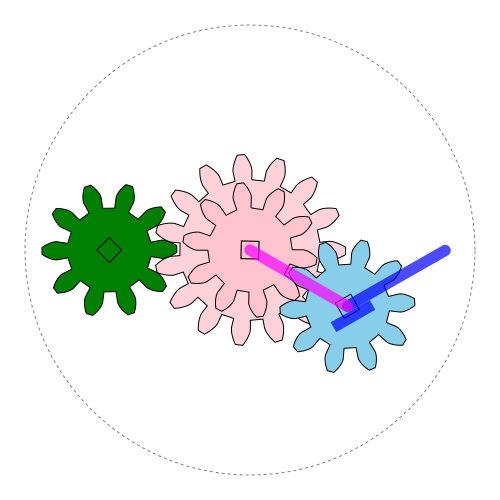}
         \caption{}
         \label{fig:orbraygears}
     \end{subfigure}%
        \caption{Trillium mechanics; (a) 3D illustration of the coupled gears; (b) Cross-section of the coupled gears.}
        \label{fig:trillium_gears}
\end{figure}

The Orbray raft is based on a design concept referred to as "\textit{Trillium}", 
which was first shown in Silber et al. (\cite{silber_25000_2022}) (Figure \ref{fig:trillium_robots}), and offered to either vendor as an open "reference" design (CAD model link: \url{https://zenodo.org/record/6354859}) as a starting point to build off. Robots are built and tested in triplets, and these triplets are inserted into the triangular module chassis. The Trillium design's alpha ($\alpha$) and beta ($\beta$) axes are mechanically coupled through a concentric gear train: rotating the alpha motor shaft while holding the beta motor shaft fixed drives the central Spur 2 cluster, causing Spur 3 (nested within the moving alpha arm) to induce an identical, parasitic rotation on the beta output axis in the same direction (see Figure \ref{fig:trillium_gears}, where the beta motor drives Spur 0 and idles on Spur 1). This coupled rotation must be compensated in software; to keep the beta output axis fixed during an alpha movement, the beta motor must be driven simultaneously by a compensating amount in the opposite direction.

The manufacturers produced their prototypes as best efforts to meet an open specification (link: \hyperlink{https://zenodo.org/records/10688871}{https://zenodo.org/records/10688871}) which we developed. We provided them with an open reference design as a starting point for the raft module as well (CAD model link: \hyperlink{https://zenodo.org/records/7226439}{https://zenodo.org/records/7226439}).

Despite the difference in mechanical architectures both module designs share the same 3-screws interface that bolts them to the focal plate, as shown in Figure \ref{fig:modular concept}. Since the robots are also driven by the same 4-mm BLDC motors the electronics interface for control is also similar. Finally, each positioner is equiped with the same optical fiber which results in the same fiber interface as well. Therefore, either design (or both) can be mounted in the same focal plane.

 \section{XY Positioning Performance Testing on Two MPS Prtotoypes}
A wide range of experimental testing has to be performed on the robotic positioners to verify their compliance with the requirements set by the projects utilizing them (\cite{kronig_optical_2020}). Previously, for surveys like SDSS-V and DESI, the robotic positioners were individual components that could be manipulated easily outside of the focal plate. However, with the modular solution, the testing requires jigs that are suitable for entire modules and not just individual positioners. Accordingly, in this paper, we have updated our testing jigs to accommodate the entire module and still be able to test the positioners one by one and later possibly all together. This section mainly focuses on the XY positioning performance testing which is a very critical test to assess the performance of the mechanics of the positioners. 

\subsection{Performance Metrics}
\label{subsec:performance_metrics}
In our evaluation, we focus on several key performance metrics. These metrics provide essential insights into the system’s behavior and guide us in optimizing its design and functionality. By meticulously analyzing these performance metrics, we gain a comprehensive understanding of the system’s strengths and areas for improvement.

\begin{itemize}

\item Repeatability
\begin{itemize}
\item Definition: Repeatability refers to the system’s ability to consistently return to the same position when repeatedly commanded to move to a specific target approaching from the same side (clockwise or counterclockwise).
\item Measurement: We perform multiple cycles of positioning the fiber to the same target and record the deviations from the expected position.
\item Desired Outcome: Lower repeatability errors indicate better precision and reliability, this allows us to have confidence in being able to calibrate the system and achieve the required positioning performance.
\end{itemize}

\item Arc Residual
\begin{itemize}
\item Definition: Arc residual refers to the error calculated when commanding the robotic arms to move small steps (typically 1 degree).
\item Measurement: Each robotic arm is commanded to move by 1 degree, and the difference between the value reached and the commanded 1 degree is considered as the error. 
\item Desired Outcome: Ideally, the arms move as closest as possible to the commanded 1 degree.
\end{itemize}

\item Datum Repeatability
\begin{itemize}
\item Definition: Datum repeatability refers to how repeatably the position at the hard stop datum is reached.
\item Measurement: Moving the positioner arms multiple times to firmly engage the arm's mechanical datums, thereby ensuring reliable datum acquisition and accurate position measurement.
\item Desired Outcome: This is the reference for any homing procedure of the system and must be repeatable. 
\end{itemize}
\item Backlash
\begin{itemize}
\item Definition: Backlash is the play or slack in the system due to mechanical clearances (e.g. gears, couplings).
\item Measurement: We measure the shift in position when going to the same target approaching from a different direction (clockwise and counterclockwise).
\item Desired Outcome: Lower dispersion in backlash values allows tighter motion path prediction for anti-collision planning.
\end{itemize}
\item Non-linearity
\begin{itemize}
\item Definition: Non-linearity assesses how closely the system’s actual movement follows a linear relationship with the commanded position.
\item Measurement: We compare the actual position (measured centroid) with the expected position at various target locations.
\item Desired Outcome: Ideally, the system should exhibit linear behavior, minimizing deviations from linearity. Due to imperfections in the gears (as shown in Figure \ref{fig:non-linearity}), there is non-linearity, but having lower values means the model can be fitted and the error is minimized.

\begin{figure}[!t]
    \centering
    \captionsetup{justification=centering}
    \includegraphics[width=0.7\linewidth]{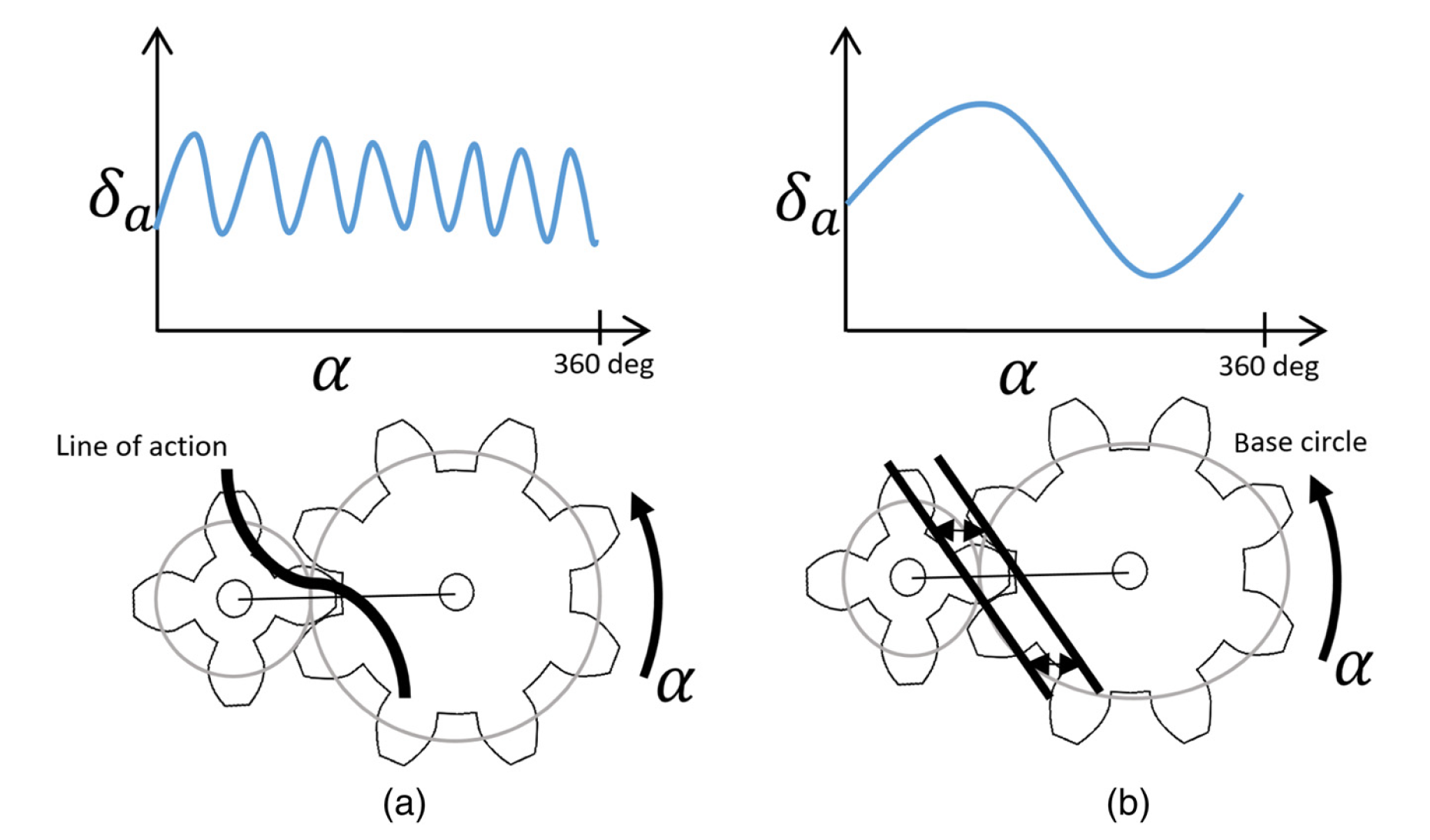}
    \caption{Non-constant reduction ratio illustrated on a simple two pinion reduction gear. (a) Illustration showing the effect of a gear tooth profile deviating from a perfect involute; (b) Illustration
showing the effect of a non-perfect base circle. Figure obtained from (\cite{kronig_precision_2020}).}
\label{fig:non-linearity}
\end{figure}

\end{itemize}
\end{itemize}

\subsubsection{XY Experimental Test-bench}
\begin{figure}[!b]
    \centering
    \includegraphics[width=0.6\linewidth]{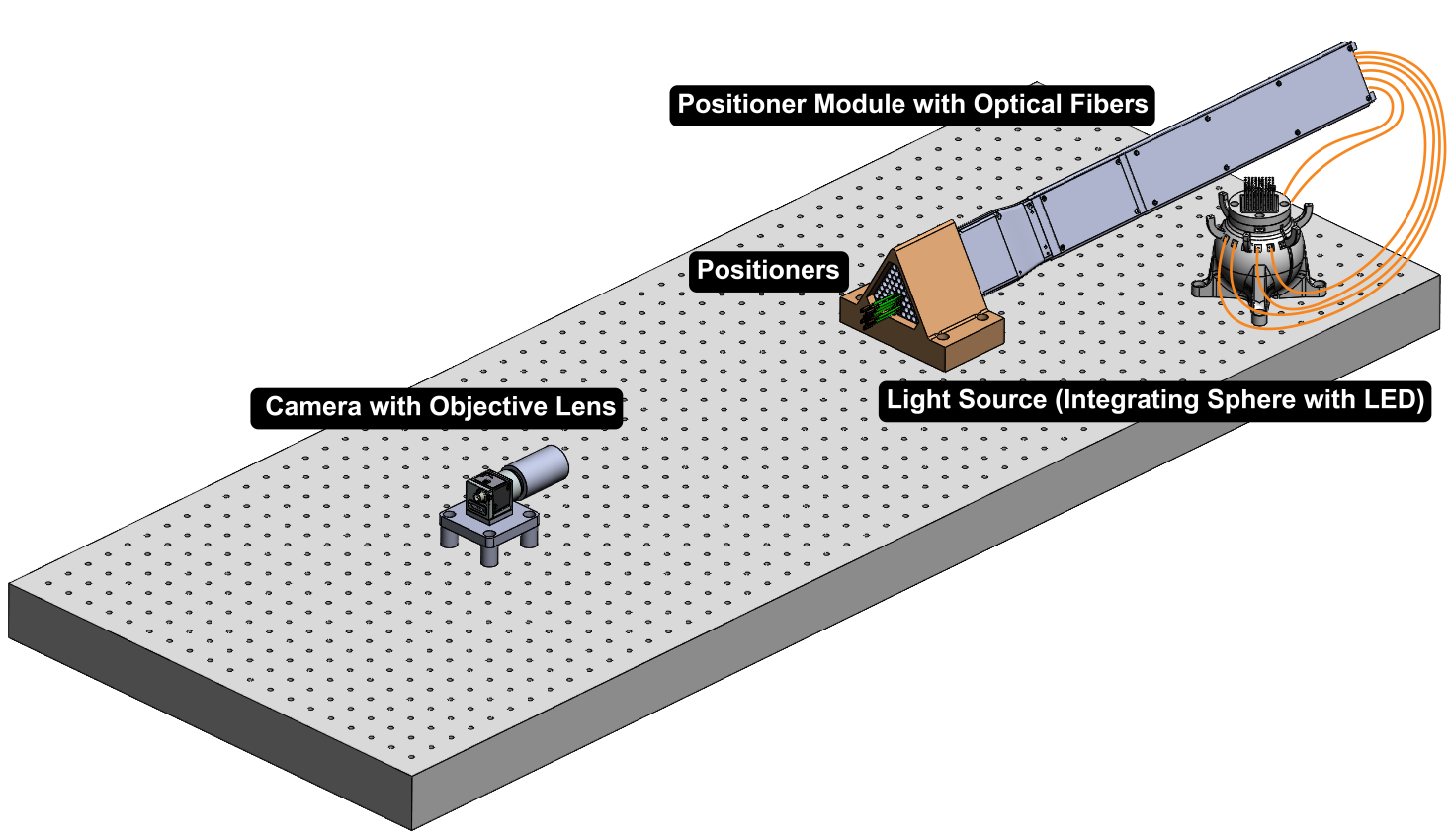}
    \caption{Rendered 3D model of the XY positioning performance test-bench.}
    \label{fig:xy_testbench}
\end{figure}
The experimental setup (shown in Figure \ref{fig:xy_testbench}) measures the x and y positions of the fibers. The fibers are backlit using a white LED as a light source illuminating the fibers. Any color for the LED source can be utilized for this setup without affecting the acquired results. If necessary, several fibers can be connected to an integrating sphere connected to an LED source to be equally and simultaneously illuminated. A Basler camera (Basler acA3800-14um) is utilized to acquire the light spot projected by the optical fiber held by the fiber positioner. For a wider field of view, another Basler camera (Basler acA5472-17um) can be used. The positioners use former SDSS-V electronics and are set to move at 1000 RPMs. Each positioner is programmed to move to given positions, and the position of the fiber is compared to the commanded position. This setup can be used for the calibration and verification of the positioner’s movement, as well as for thermal and lifetime tests.

\subsection{XY Test Results}

\begin{figure}[!t]
    \centering
    \captionsetup[subfigure]{justification=centering}
    \begin{subfigure}[b]{0.33\textwidth}
    \centering
    \includegraphics[height=4cm,trim={1cm 1cm 1cm 1cm},clip]{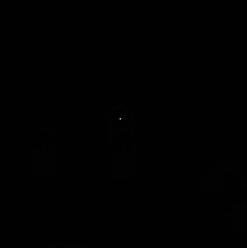}
    \caption{}
    \label{fig:blob_zoomedout}
\end{subfigure}
\hspace{1mm}
\begin{subfigure}[b]{0.33\textwidth}
    \centering
    \includegraphics[height=4cm,trim={1cm 1cm 1cm 1cm},clip]{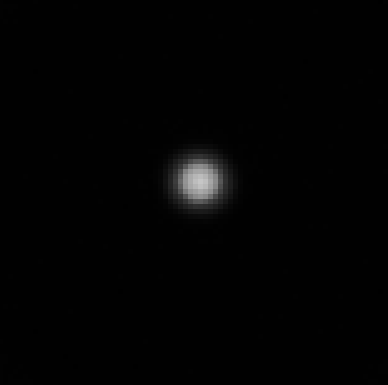}
    \caption{}
    \label{fig:blob_zoomedin}
\end{subfigure}
    \caption{Illuminated spot captured by the camera; (a) Original image of what the camera sees showing a very small dot in the middle of the image; (b) Zoomed-in version of (a).}
    \label{fig:XY_light_blob}
\end{figure}

The calculation of the XY results is made by acquiring illuminated spots and calculating their centroids to retrieve their XY coordinates. An example of such illuminated spots is found in Figure \ref{fig:XY_light_blob}.

\subsubsection{XY Positioning Repeatability}
\begin{figure}[!b]
     \centering
     \captionsetup[subfigure]{justification=centering}
     \begin{subfigure}[b]{0.5\textwidth}
         \centering
         \includegraphics[width=\textwidth]{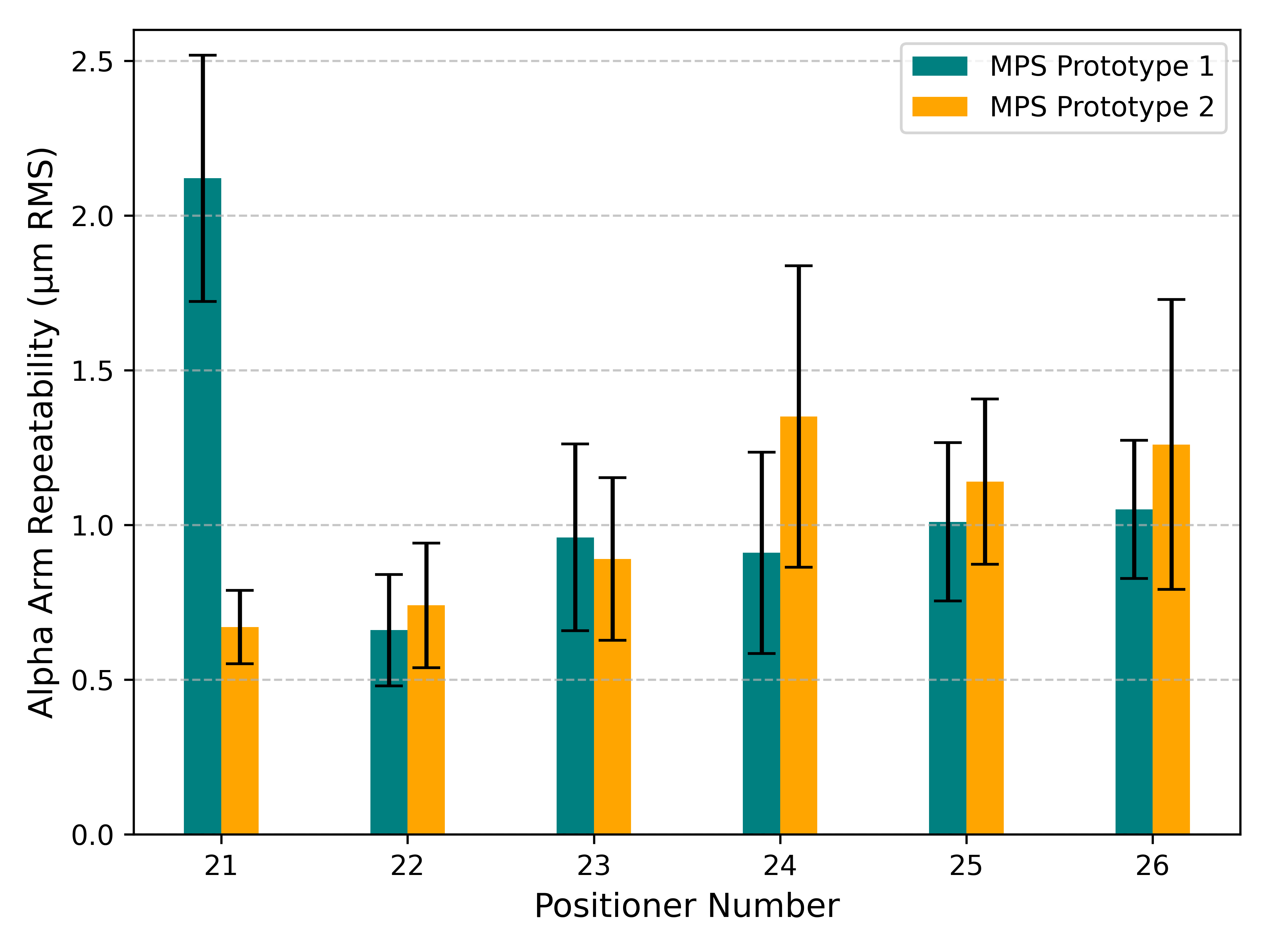}
         \caption{Alpha Arm}
         \label{fig: alpha_arm_repeatability}
     \end{subfigure}%
     \hfill
     \begin{subfigure}[b]{0.5\textwidth}
         \centering
         \includegraphics[width=\textwidth]{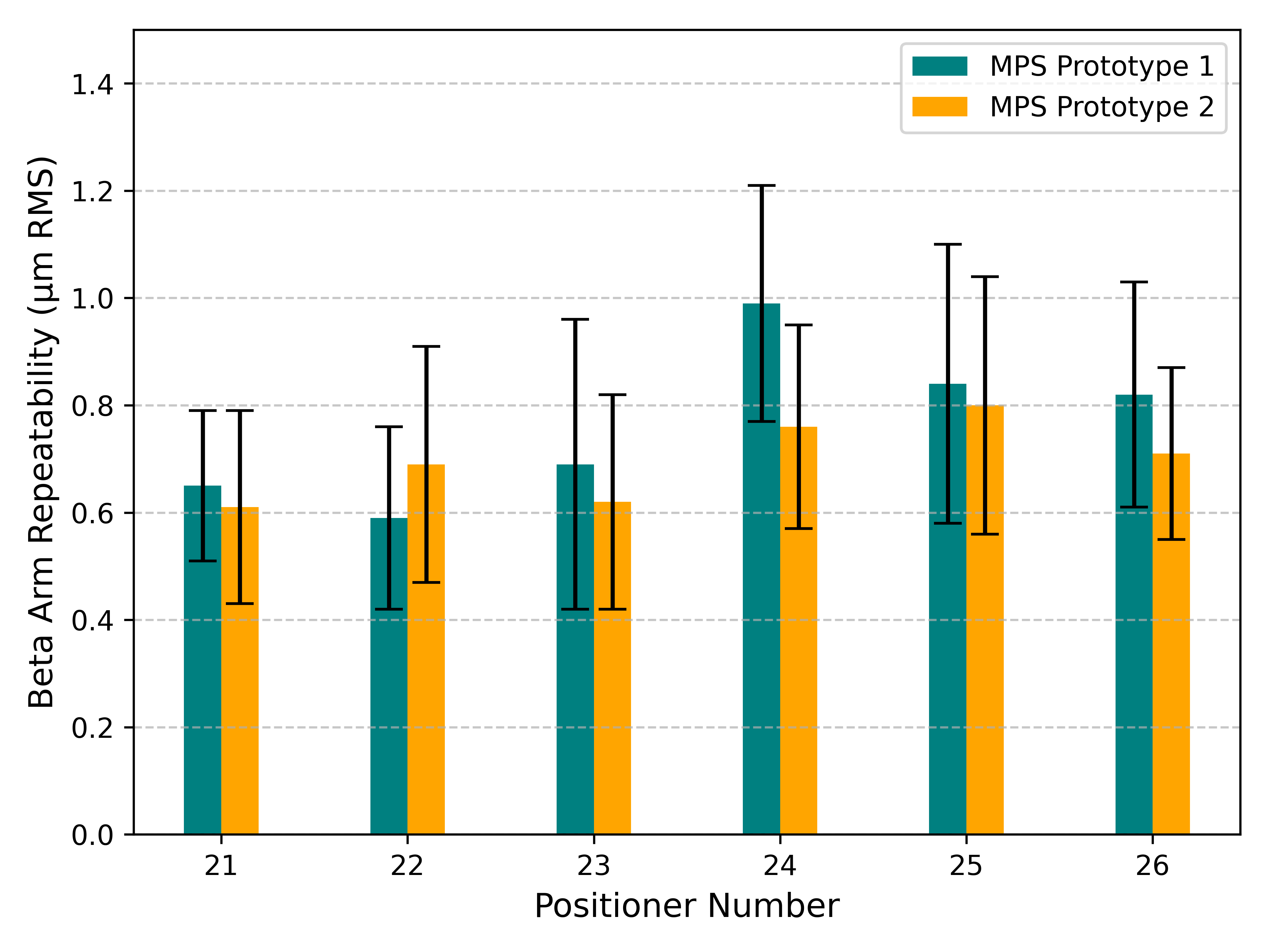}
         \caption{Beta Arm}
         \label{fig: beta_arm_repeatability}
     \end{subfigure}%
        \caption{Graphs showing the \textbf{XY positioning performance repeatability} for two MPS prototypes. The bar plots show the root-mean-square values versus the positioner number and the standard deviation of each bar; (a) results for the alpha arm, and (b) results for the beta arm.}
        \label{fig: x-y repeatability}
\end{figure}
For the positioning repeatability measurement, the data is acquired by moving the alpha arm and beta arms to particular positions multiple times. The centroids of the illuminated spots are found using 2D Gaussian fitting, and the root mean square of the repeatability is calculated from 20 iterations of reaching each XY position. The data presented in Figure \ref{fig: x-y repeatability} shows the mean of the root mean square values of the program repeated 24 times.

\subsubsection{Datum Repeatability}
\begin{figure}[!t]
     \centering
     \captionsetup[subfigure]{justification=centering}
     \begin{subfigure}[b]{0.5\textwidth}
         \centering
         \includegraphics[width=\textwidth]{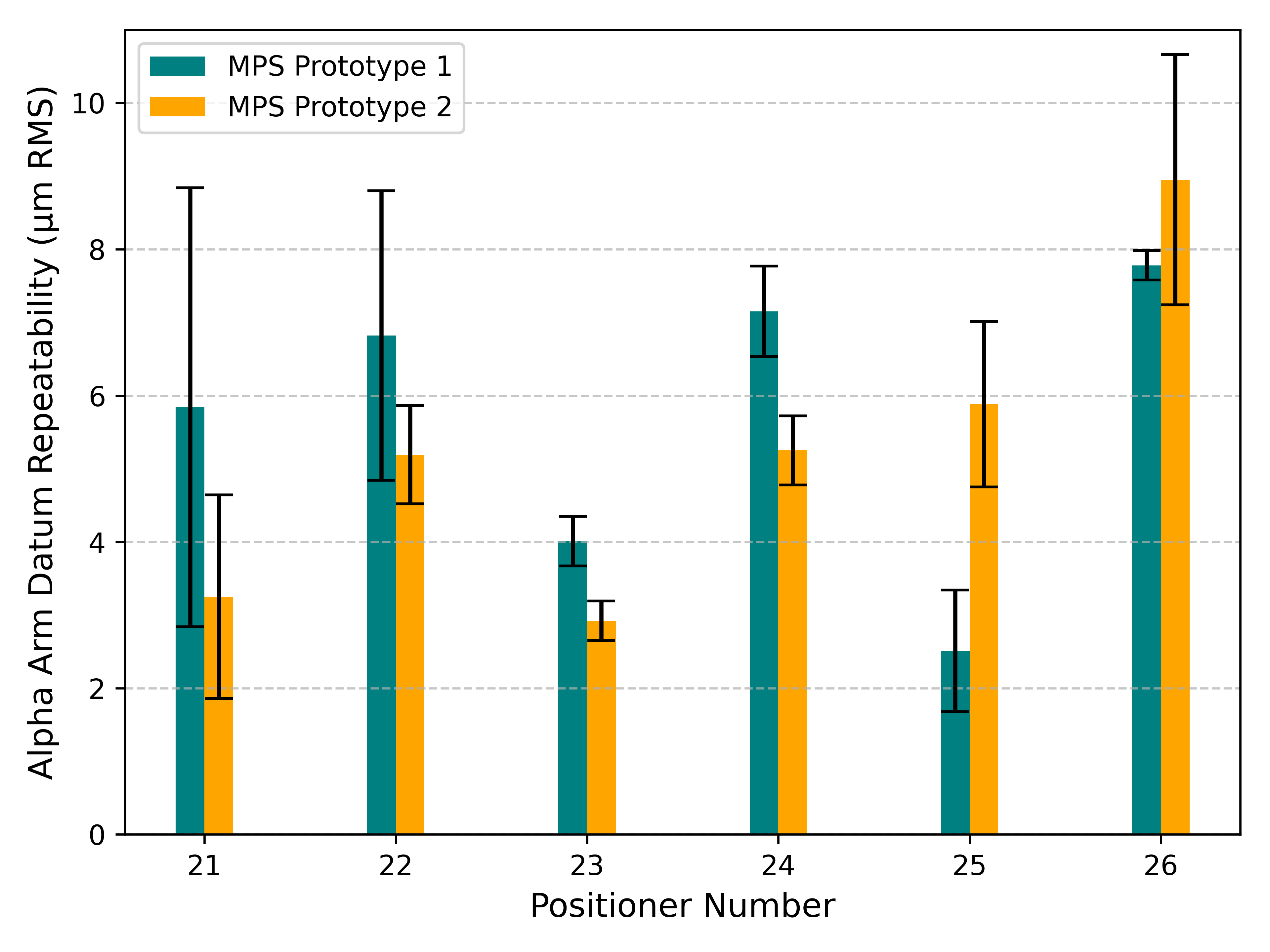}
         \caption{Alpha Arm}
         \label{fig: alpha_datum}
     \end{subfigure}%
     \hfill
     \begin{subfigure}[b]{0.5\textwidth}
         \centering
         \includegraphics[width=\textwidth]{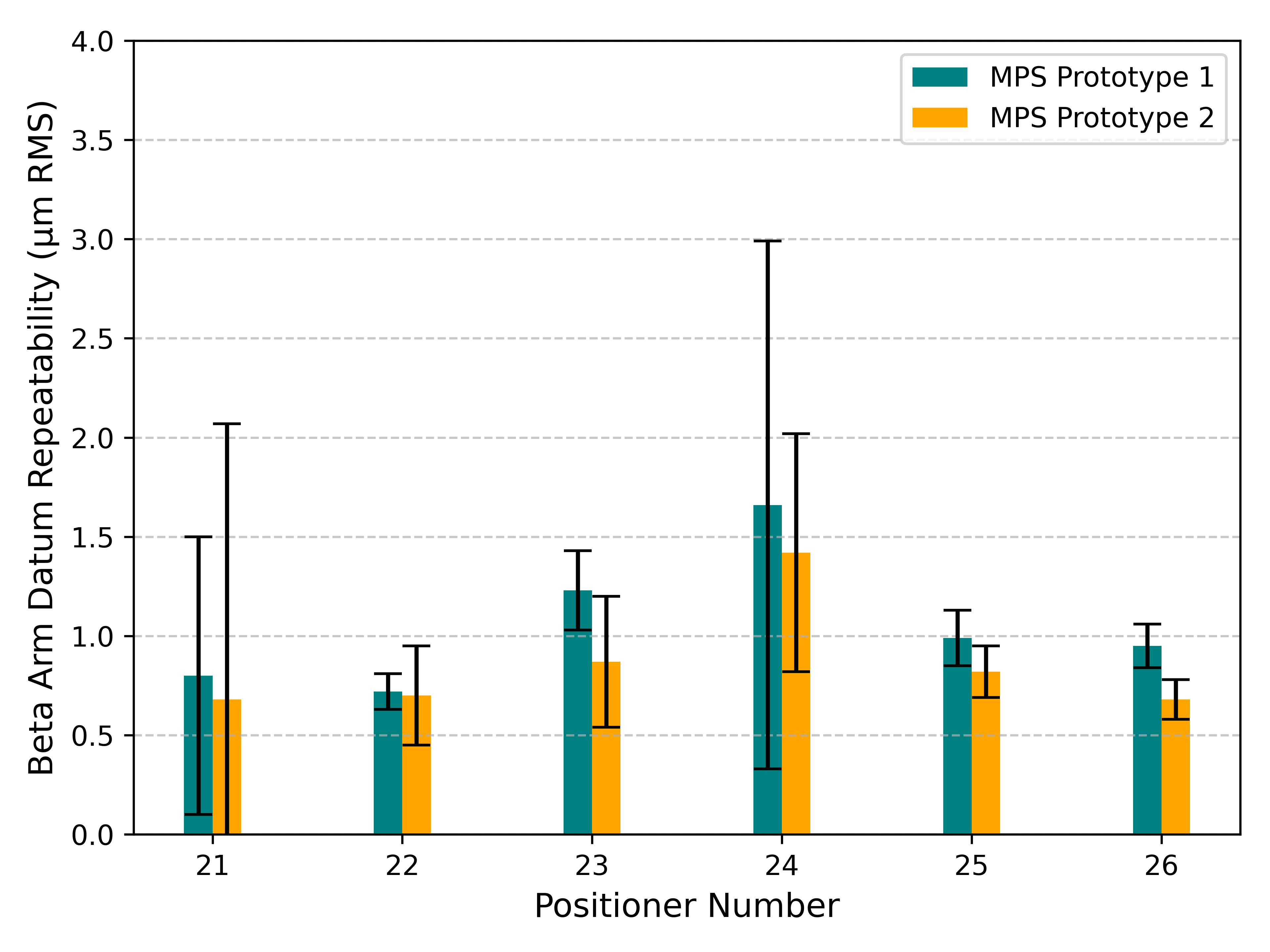}
         \caption{Beta Arm}
         \label{fig: beta_datum}
     \end{subfigure}%
        \caption{Graphs showing the \textbf{datum repeatability} for two MPS prototypes. The bar plots show the root-mean-square values versus the positioner number and the standard deviation of each bar; (a) results for the alpha arm, and (b) results for the beta arm.}
        \label{fig: Datum repeatability}
\end{figure}
To assess the datum (hard stop) repeatability, the data is acquired by moving the alpha and beta arms to their hard stops. The centroids of the illuminated spots are found using 2D Gaussian fitting, and the root mean square of the repeatability is calculated from 100 iterations of reaching each XY position. The data in Figure \ref{fig: Datum repeatability} shows the mean of the root mean square values of the program repeated 24 times.

Subsequent discussions on testing methodology highlighted an additional effect relevant to datum repeatability: gearbox stress loading at the datum. The robots are positioned using Field-Oriented Control (FOC) of their BLDC motors. Owing to tight space and cost constraints, the motors are operated in a sensorless configuration, without Hall sensors or encoders. Accurate knowledge of the datum position can be helpful in practice during integration and test: the datum defines an origin reference for the positioner axes even in cases where the field survey camera may not yet be installed or is otherwise offline.

In the current control scheme, transitions between angular positions are achieved by energizing the motors for a predetermined duration, based on the estimated current position and the gearbox ratio. As a result, when the mechanism reaches the datum, the commanded rotation may slightly exceed the required angular travel. In such cases, the rotating stator magnetic field continues to apply torque to the rotor, leading to stress loading of the gearbox against the mechanical stop. The resulting preload depends on the mechanical stiffness, compliance, and friction, and may influence the achieved datum position.

While this effect was not explicitly isolated in the present test campaign, it has been identified as a relevant aspect for further investigation and will be addressed in future testing. Importantly, the datum repeatability results presented here, obtained during early prototyping, remain representative of the expected order of magnitude and provide a meaningful baseline for subsequent design and control refinements.

\subsubsection{Backlash}
\label{subsec:backlash}
\begin{figure}[!t]
     \centering
     \captionsetup[subfigure]{justification=centering}
     \begin{subfigure}[b]{0.5\textwidth}
         \centering
         \includegraphics[width=\textwidth]{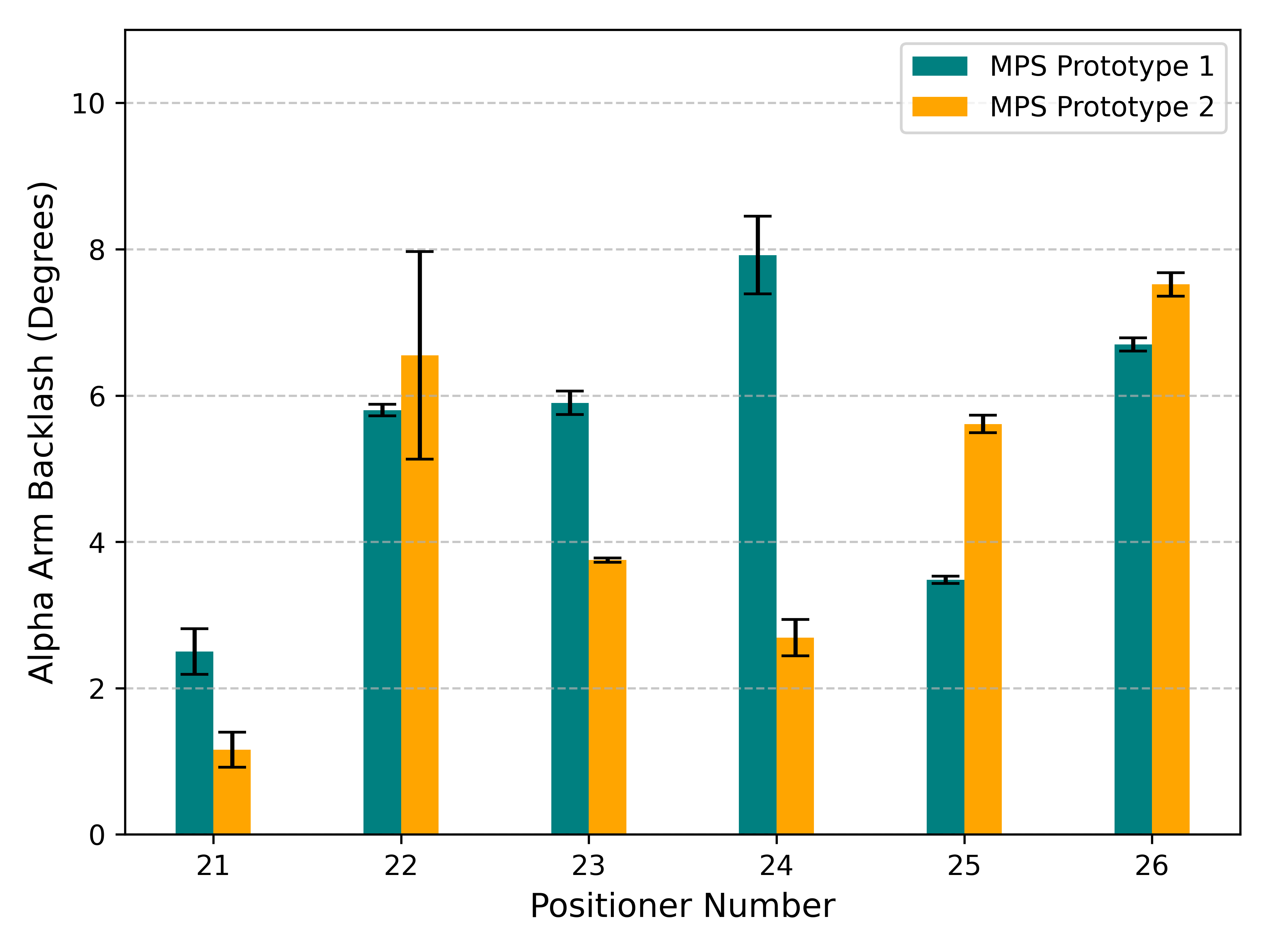}
         \caption{Alpha Arm}
         \label{fig: alpha_backlash}
     \end{subfigure}%
     \hfill
     \begin{subfigure}[b]{0.5\textwidth}
         \centering
         \includegraphics[width=\textwidth]{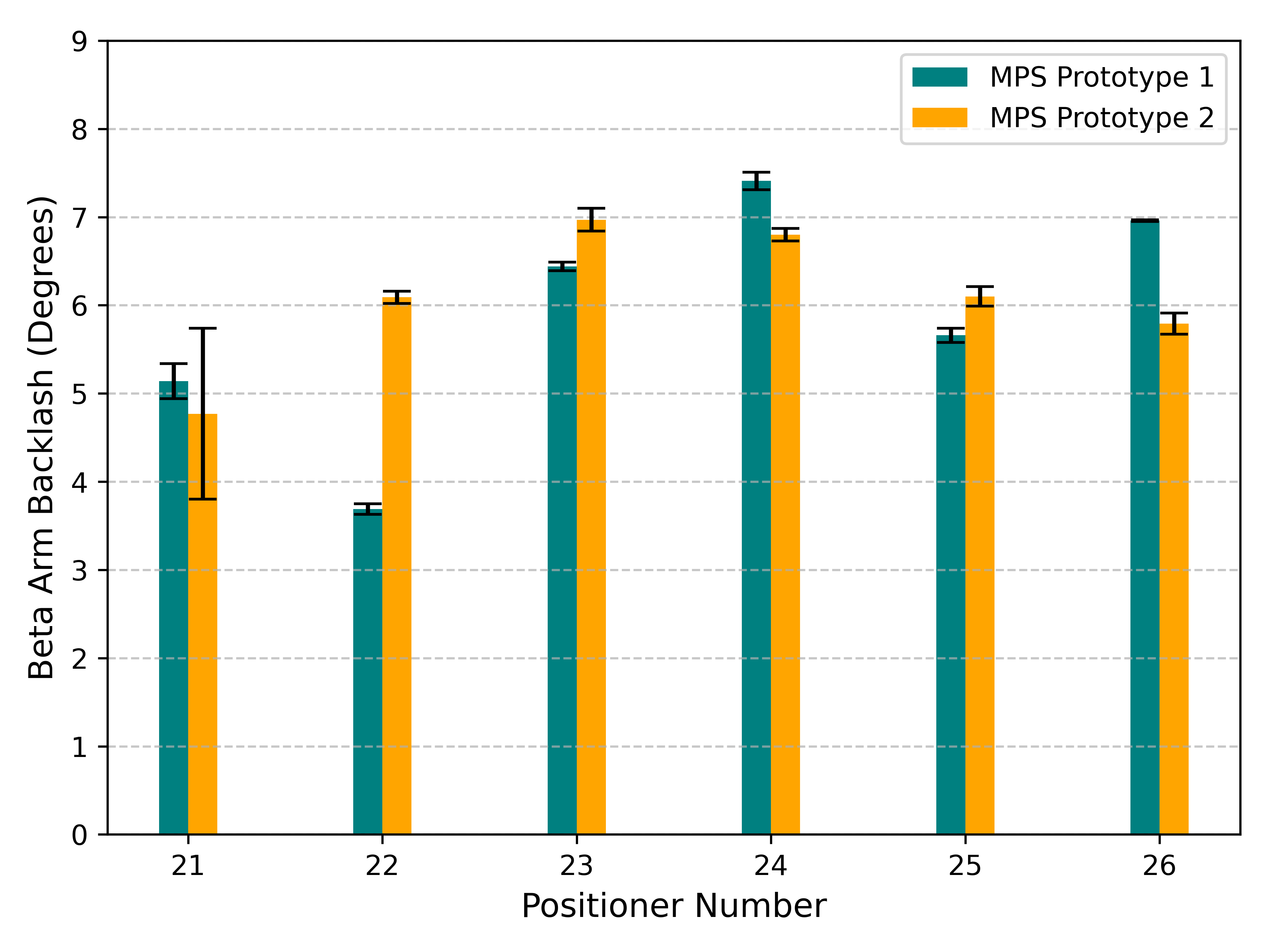}
         \caption{Beta Arm}
         \label{fig: beta_backlash}
     \end{subfigure}%
        \caption{Graphs showing the \textbf{backlash} for two MPS prototypes. The bar plots show the root-mean-square values versus the positioner number and the standard deviation of each bar; (a) results for the alpha arm, and (b) results for the beta arm.}
        \label{fig: Backlash}
\end{figure}
For the backlash measurement, the positioner is commanded to move to a particular position from the two directions and the centroids of the illuminated spots are calculated for each direction. The angle between the two centroids for each direction is the backlash. The data shown in Figure \ref{fig: Backlash} presents the mean of the root mean square values of the program repeated 24 times. 

\begin{figure}[!b]
     \centering
     \captionsetup[subfigure]{justification=centering}
     \begin{subfigure}[b]{0.35\textwidth}
         \centering
         \includegraphics[width=\textwidth]{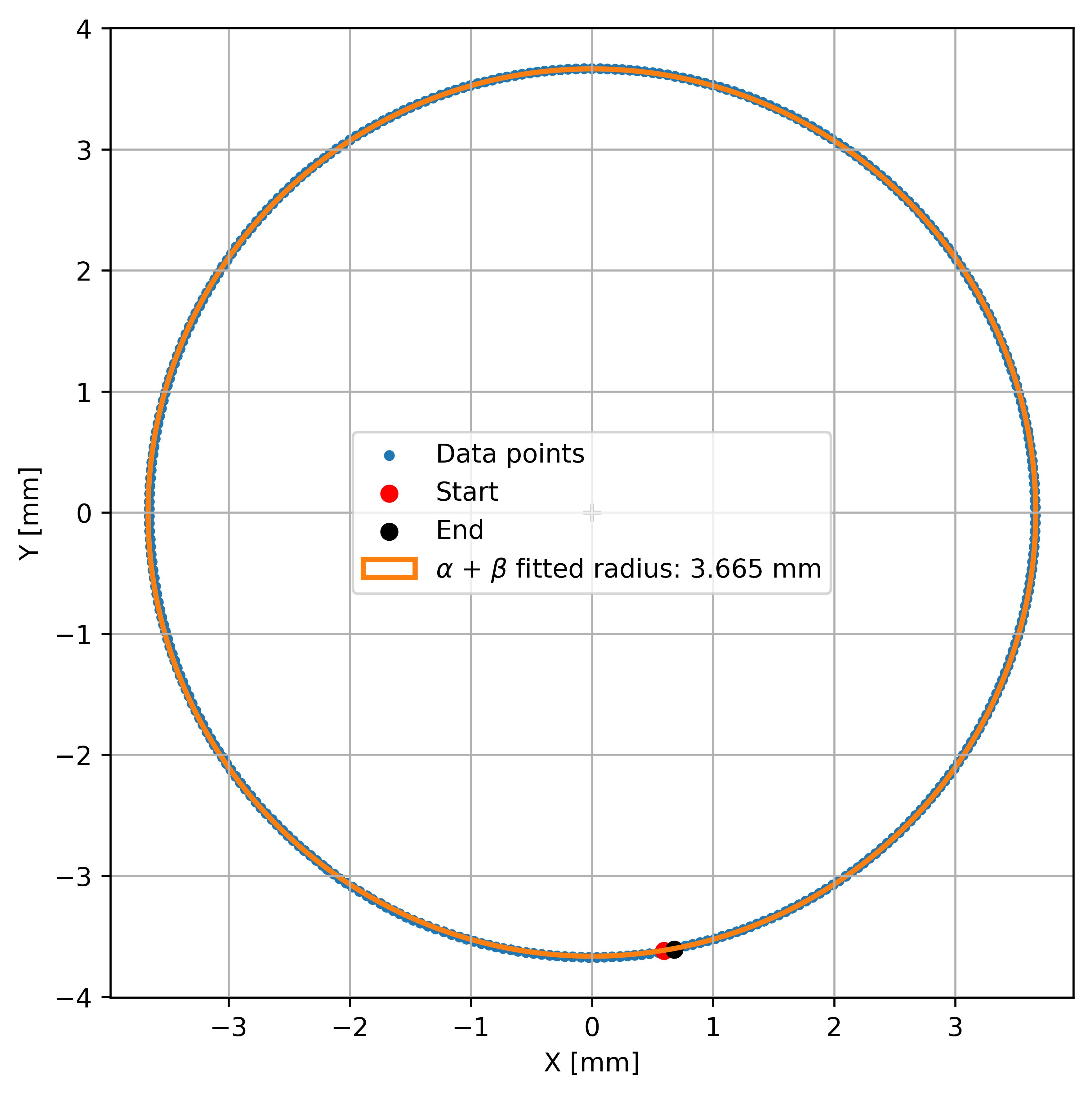}
         \caption{}
         \label{fig: alpha_nl_circle}
     \end{subfigure}%
     \hspace{0.5cm}
     \begin{subfigure}[b]{0.35\textwidth}
         \centering
         \includegraphics[width=\textwidth]{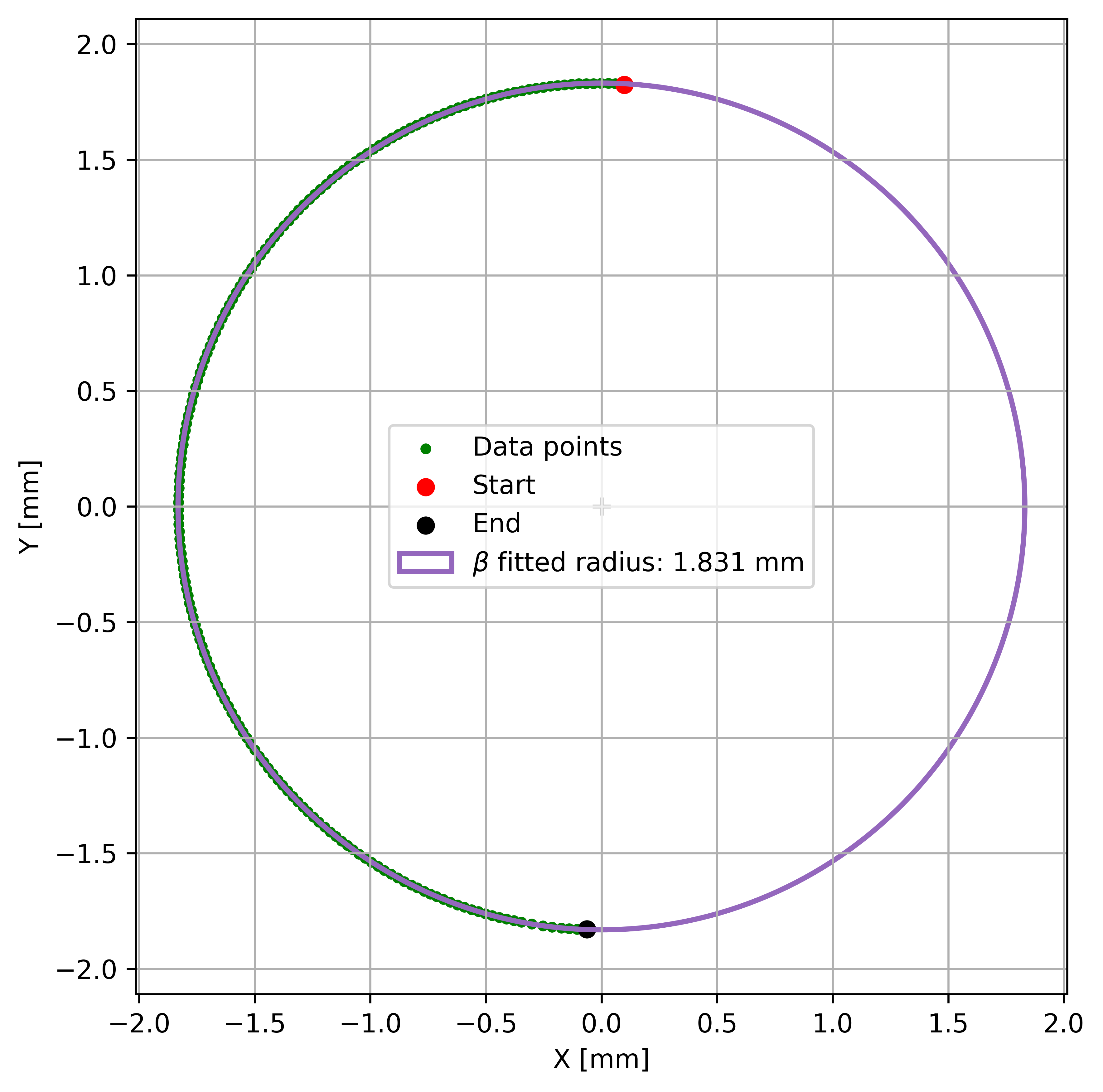}
         \caption{}
         \label{fig: beta_nl_arc}
     \end{subfigure}%
        \caption{(a) Graph showing the circle constructed by rotating the alpha arm in steps of 1° with the beta arm fixed at 180°; (b) Graph showing the arc constructed by rotating the beta arm in steps of 1° with the alpha arm fixed at 0°.}
        \label{fig: nl_circle_arc}
\end{figure}

\begin{figure}[!t]
    \centering
    \captionsetup{justification=centering}
    \includegraphics[width=0.8\linewidth]{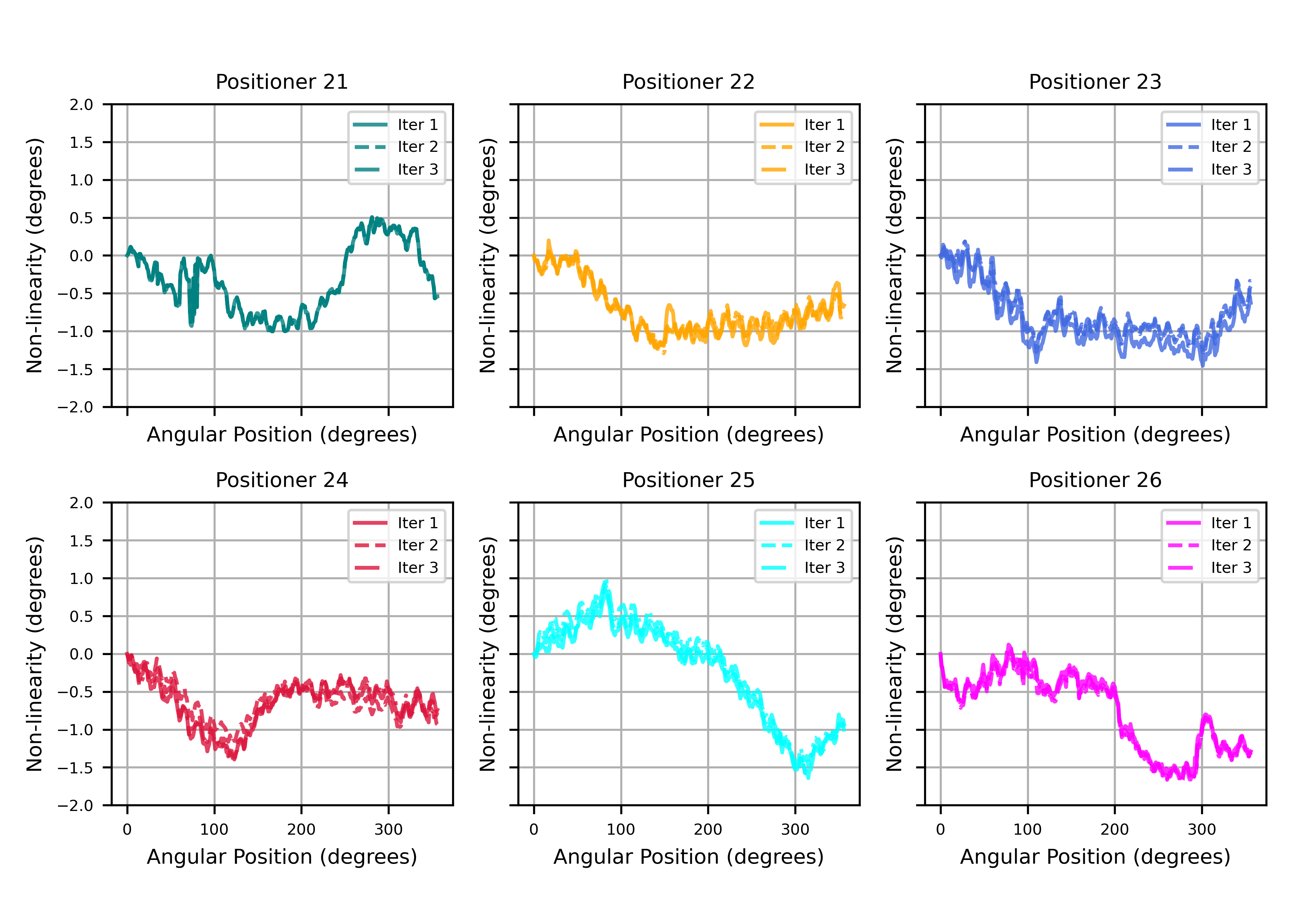}
    \caption{Graphs showing the \textbf{non-linearity} curves of MPS Prototype 1 for the \textit{alpha} arm of all positioners.}
\label{fig: alpha non-linearity}
\end{figure}

\begin{figure}[!t]
    \centering
    \captionsetup{justification=centering}
    \includegraphics[width=0.8\linewidth]{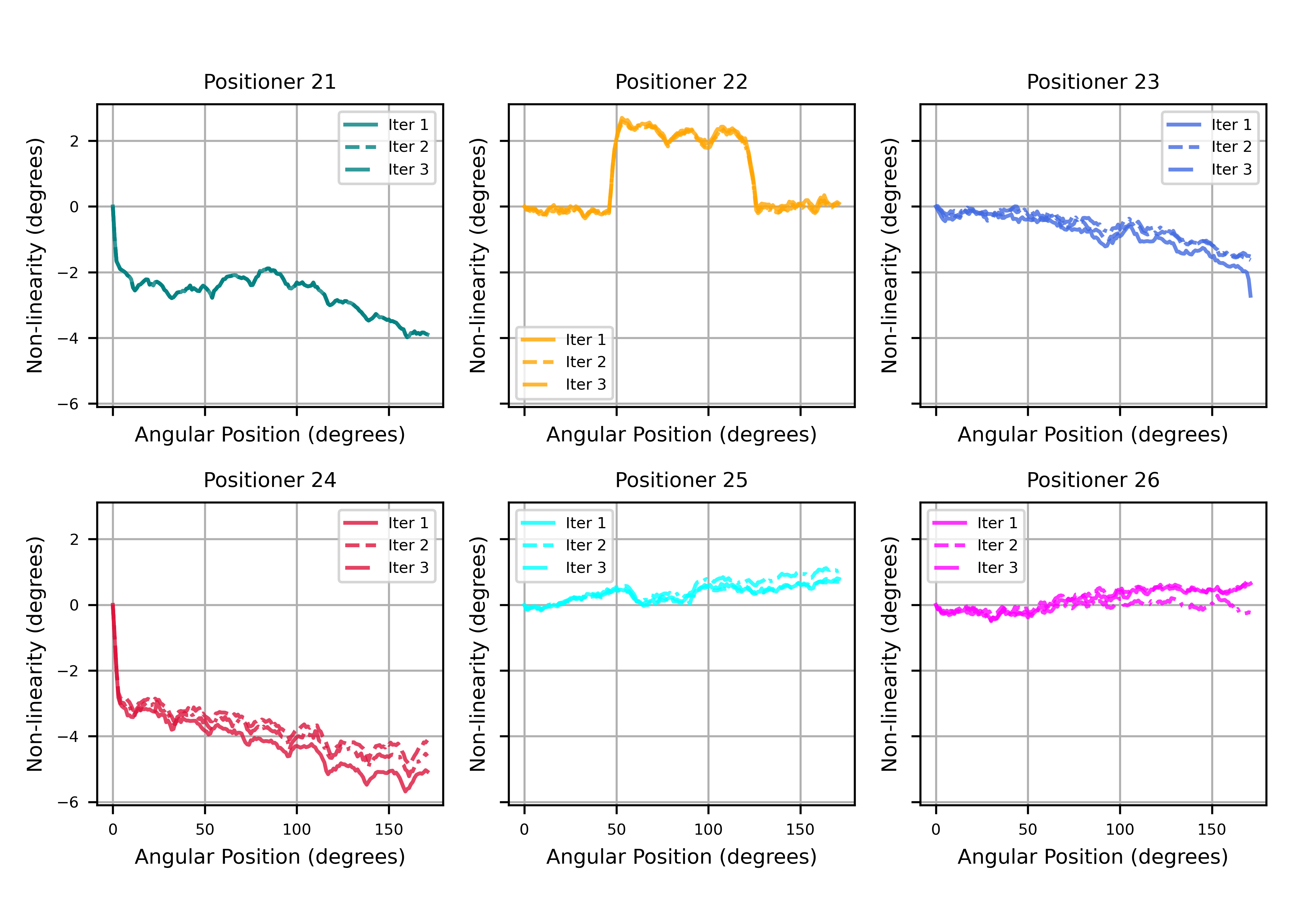}
    \caption{Graphs showing the \textbf{non-linearity} curves of MPS Prototype 2 for the \textit{beta} arm of all positioners.}
\label{fig: beta non-linearity}
\end{figure}

\subsubsection{Non-linearity}
\label{subsec: non-linearity}
The non-linearity measurement is conducted by commanding the positioner to move by one degree over the full angular range of both alpha and beta arms. As shown in Figure \ref{fig: alpha_nl_circle}, the data points result in a circle covering the entire angular range of the alpha arm. It is constructed by rotating the alpha arm in steps of 1° with the beta arm fixed at 180°. The data points are fitted with a circle whose radius is measured to be about 3.665 mm. The same applies for the beta arm which is rotated in steps of 1° with the alpha arm fixed at 0°. The data points then construct an arc which is fitted by a circle (shown in Figure \ref{fig: beta_nl_arc}) whose radius is 1.831 mm. The radius of the beta arm directly infers the arm length which is very close to the theoretical CAD value of 1.8 mm. The arm length of the alpha arm can be calculated by subtracting the length of the beta arm from the radius of the constructed alpha circle, and that gives 1.834 mm (for reference, the theoretical CAD value is 1.8 mm). 

\begin{table}[!t]
    \centering
    \small
    \renewcommand{\arraystretch}{1.3}
    \begin{tabular}{|c|*4{w{c}{1.8cm}|}} \hline
        \multirow{3}{*}{\shortstack{Positioner \\ ID}} & \multicolumn{2}{c|}{\parbox[t]{4cm}{\centering Arc Residual [$\mu m$] \\ (MPS Prototype 1)}} & \multicolumn{2}{c|}{\parbox[t]{4cm}{\centering Arc Residual [$\mu m$] \\ (MPS Prototype 2)}} \\ \cline{2-5}
          & Alpha & Beta & Alpha & Beta \\ \hline
         21 & 5.83 & 2.08 & 3.39 & 2.61 \\ \hline
         22 & 2.91 & 3.99 & 3.63 & 1.61 \\ \hline 
         23 & 3.41 & 3.33 & 3.29 & 2.54 \\ \hline 
         24 & 2.67 & 2.38 & 2.35 & 1.49 \\ \hline 
         25 & 2.89 & 1.52 & 3.59 & 1.29  \\ \hline 
         26 & 3.70 & 1.99 & 2.50 & 2.17  \\ \hline
    \end{tabular}
    \vspace{5pt}
    \caption{The arc residuals are calculated as the target angle minus the measured angle over 1 degree moves that span the range of each axis. The arc residual results shown in the table are for two MPS prototypes tested at EPFL.}
    \label{table:MPS_arc_residual}
\end{table}

Figures \ref{fig: alpha non-linearity} and \ref{fig: beta non-linearity} show examples of the non-linear behavior of the alpha and beta arms, respectively, for the MPS Prototype 1. The high-frequency signal of the transmission error ($\delta_a$) shown in the figures is the result of a deviation from a perfect involute tooth profile causing a curved line of action instead of a straight one, as explained previously by Figure \ref{fig:non-linearity} (\cite{kronig_precision_2020}). It is suspected that those high frequency peaks correspond to the number of teeth of the last transmission stage gear linked to the given axis of rotation; alpha for Figure \ref{fig: alpha non-linearity} and beta for Figure \ref{fig: beta non-linearity}. An example of an apparent low-frequency signal from the transmission error ($\delta_a$) can be seen in Figure \ref{fig: beta non-linearity} for positioner 22. This is likely caused by an imperfect base circle, possibly due to run-out, eccentricity, or mounting misalignment (\cite{kronig_precision_2020}).

Similarly to how the non-linearity is being measured, using the 1 degree movement of the robotic arms, the arc residual has been estimated for the small moves of the arms. The results of the small move error are presented in Table \ref{table:MPS_arc_residual} for both MPS prototypes tested at EPFL. The error is estimated in degrees and is then converted to $\mu m$ RMS by multiplying the degree value by the arm length; 1.8 mm for the beta arm, and 3.6 mm for the alpha arm.

\section{Angular Tilt Testing on One MPS Prototype}
Angular tilt testing is a vital step to ensure the highest performance of the positioner not only mechanically but also optically. Large tilts in the positioner will result in severely degraded optical performance due to Focal Ratio Degradation (FRD) which cannot be easily corrected after having already been installed on the telescope. Therefore, prior tilt testing is absolutely necessary to be able to further improve the prototyping and make sure that the final product exhibits the best performances.
\subsection{Angular Tilt Axes} \label{sec:angular_tilt_axes}

\begin{figure*}[b!]
    \centering
    \includegraphics[width=0.7\textwidth]{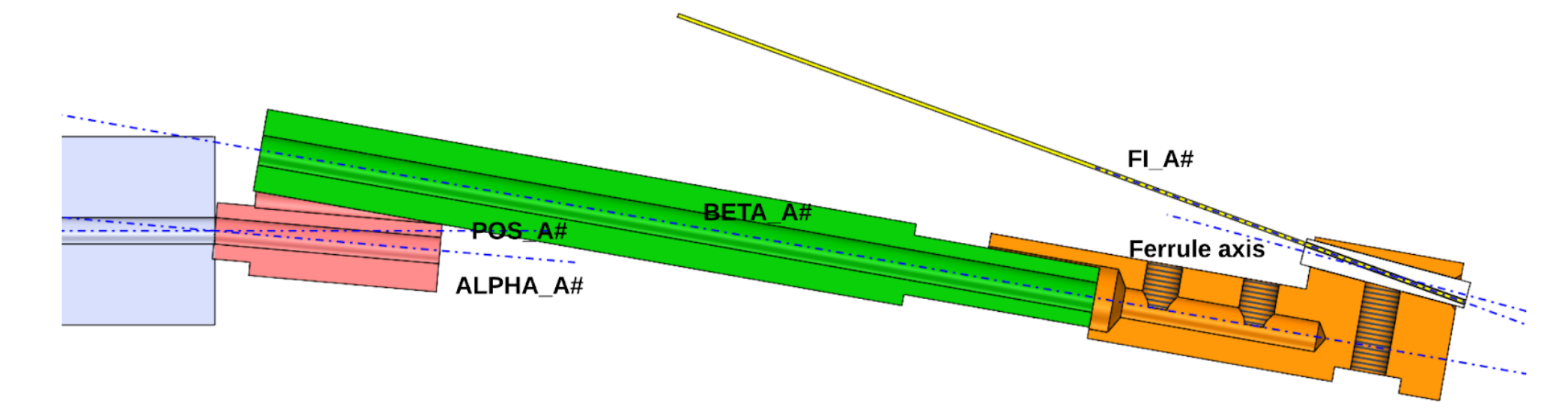}
    \caption{Schematic showing the different tilt axes.}
    \label{fig:tiltaxes}
\end{figure*}

Figure \ref{fig:tiltaxes} shows the different tilt axes that need to be considered for the robotic positioner inside the module. Table \ref{tab:tilt_axes} explains in more detail the different axes. The requirement set by several stage-5 surveys is for the overall tilt from the fiber to the module axis to be < 0.5°. It should be noted that the contribution of the tilt between the $Pos_A\#$ and the $Mod_A\#$ is not presented in this work, as it requires additional investigations and considerations. It has, however, been estimated from the worst-case scenario of the manufacturing tolerances to be about 0.05° which can be considered a negligible contribution compared to the main contributors presented here.\\
\renewcommand{\arraystretch}{1.4}
\begin{table}[ht]
\centering
\begin{tabular}{ll}
\toprule
\textbf{Tilt Axes} & \textbf{Description} \\
\midrule
$FI_A\#$ and Ferrule Axis & Tilt between fiber and ferrule axes \\
Ferrule Axis and $Beta_A\#$ & Tilt between ferrule and beta arm axes \\
$Beta_A\#$ and $Alpha_A\#$ & Tilt between beta arm and alpha arm axes \\
$Alpha_A\#$ and $Pos_A\#$ & Tilt between alpha arm and positioner axes \\
$Pos_A\#$ and $Mod_A\#$ & Tilt between positioner and module axes \\
\bottomrule
\end{tabular}
\caption{Description of the different tilt axes of the module}
\label{tab:tilt_axes}
\end{table}

\subsection{Tilt Test Concept}
\begin{figure*}[t!]
    \centering
    \includegraphics[width=0.6\textwidth]{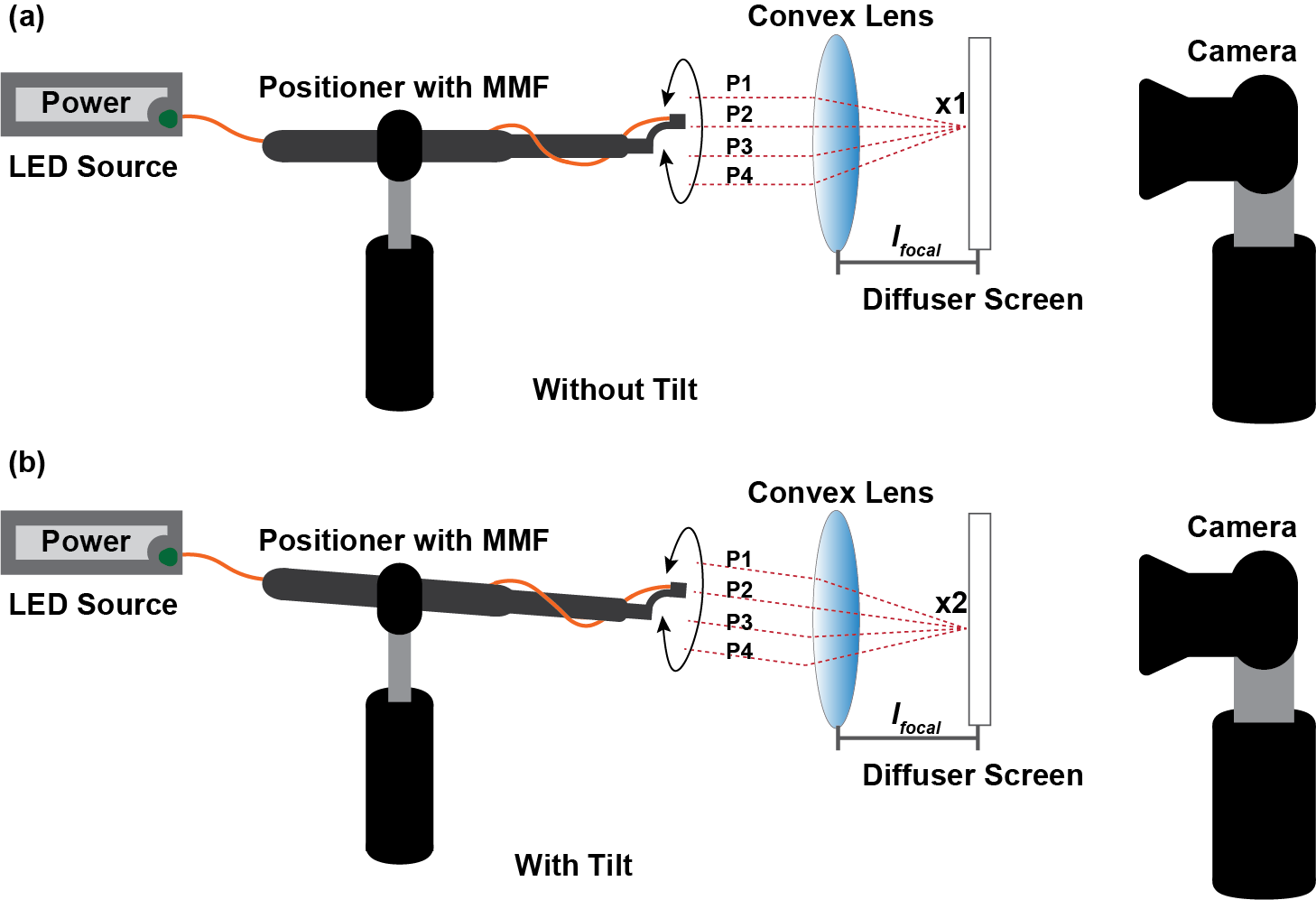}
    \caption{Schematic showing the angular tilt test concept with an LED source connected to a multi-mode fiber (MMF) held by a robotic positioner; P1-P4 are different XY positions; f is the focal length of the bi-convex lens; (a) scenario without tilt; (b) scenario with tilt.}
    \label{fig:tiltconcept}
\end{figure*}
The tilt test concept (\cite{kronig_optical_2020, kronig_precision_2020}) heavily relies on the disentanglement of the XY movement and the tilt movement. The experimental configuration shown in Figure \ref{fig:tiltconcept} using the bi-convex lens and the diffuser screen positioned at its focal length is what gives the possibility to measure tilt without any influence from the XY translation. Figure \ref{fig:tiltconcept}(a) shows an LED source connected to a multi-mode fiber (MMF) which is held by a robotic positioner. The positioner translates in x and y reaching the positions P1 to P4. It should be noted that what we see in the figure are only the principle rays of the diverging beam at each XY position, and the rest of the light rays in the beam are not shown. Irrespective of the XY movement, the point $x_1$ on the diffuser screen is almost stationary. However, if there is any tilt on the positioner's side (tilt in fiber or tilt in the positioner's arms), the principle rays will tilt accordingly, and a new point $x_2$ will be created. The displacement $\Delta d$ corresponds to the difference of $x_2$ and $x_1$ and can be calculated using the following equation:
\begin{equation}
    \Delta d = x_2 - x_1
\end{equation}
Accordingly, the tilt angle can be calculated using the following formula:
\begin{equation}
    \omega_{tilt} = tan^{-1}\left(\frac{\Delta d}{l_{focal}}\right)
\end{equation}
where $\omega_{tilt}$ is the tilt angle and $l_{focal}$ is the focal length of the convex lens.

\subsection{Disentanglement Verification Test}
The resolution of our system is defined by our ability to cleanly 
decouple translational motion (XY) from angular motion (tip-tilt). In an ideal 
measurement setup, pure translation of the module parallel to the optical axis 
should result in zero displacement of the illuminated spot on the far-field 
diffuser screen. However, due to finite alignment tolerances and system resolution 
constraints, a physical translation can project a minor apparent shift onto the 
screen. To isolate and normalize our true tilt measurements, this baseline 
translational cross-contamination must be quantified and subtracted.

To measure this resolution limit, the module is mounted on a combined translation 
and tip-tilt stage assembly. The diffuser screen is placed at the exact focal length of the lens to act as a far-field angular mapper. During this calibration step, the translation stage is moved by a known physical distance, and any minor displacement of the illuminated spot on the screen is recorded. This baseline apparent tilt is then mapped to the physical translation to determine the normalization factor.

In our current setup, a $1^\circ$ movement on the tilt stage corresponds to a 
physical displacement of the cylinder by $2\text{ mm}$. Conversely, translating the stage by $2\text{ mm}$ results in a minor shift of the illuminated spot on the screen by $3\text{ pixels}$. Using a calibration factor of $0.0741\text{mm/pixel}$, this $3\text{-pixel}$ shift equates to $0.2223\text{ mm}$. Utilizing the lens focal length ($f = 600\text{ mm}$), the angular resolution floor used to normalize and correct our final tilt data is computed via the arctangent function:

\begin{equation}
\Delta\theta = \arctan\left(\frac{0.2223\text{ mm}}{600\text{ mm}}\right) \approx 0.0212^\circ
\end{equation}

\subsection{Angular Tilt Experimental Test-bench}
For the angular tilt testing, we have implemented two different configurations and have verified their equivalence. The first test-bench presented in Section \ref{sec5: short FL setup} is based on the work by Kronig et al. (\cite{kronig_optical_2020, kronig_precision_2020}). We have used this setup with a short focal length lens (100 mm), and when we wanted to increase the system resolution, we realized that a longer focal length would mean requiring a longer optical table to conduct the experiment properly. This, in turn, resulted in slightly modifying the original setup to accommodate the longer focal length but still fit on the table. The modified long-focal length test-bench is presented in Section \ref{sec5: long FL setup}. 
\subsubsection{Short Focal Length Test-bench (Transmission Setup)}
\label{sec5: short FL setup}
\begin{figure}[!t]
    \centering
    \includegraphics[width=0.6\linewidth]{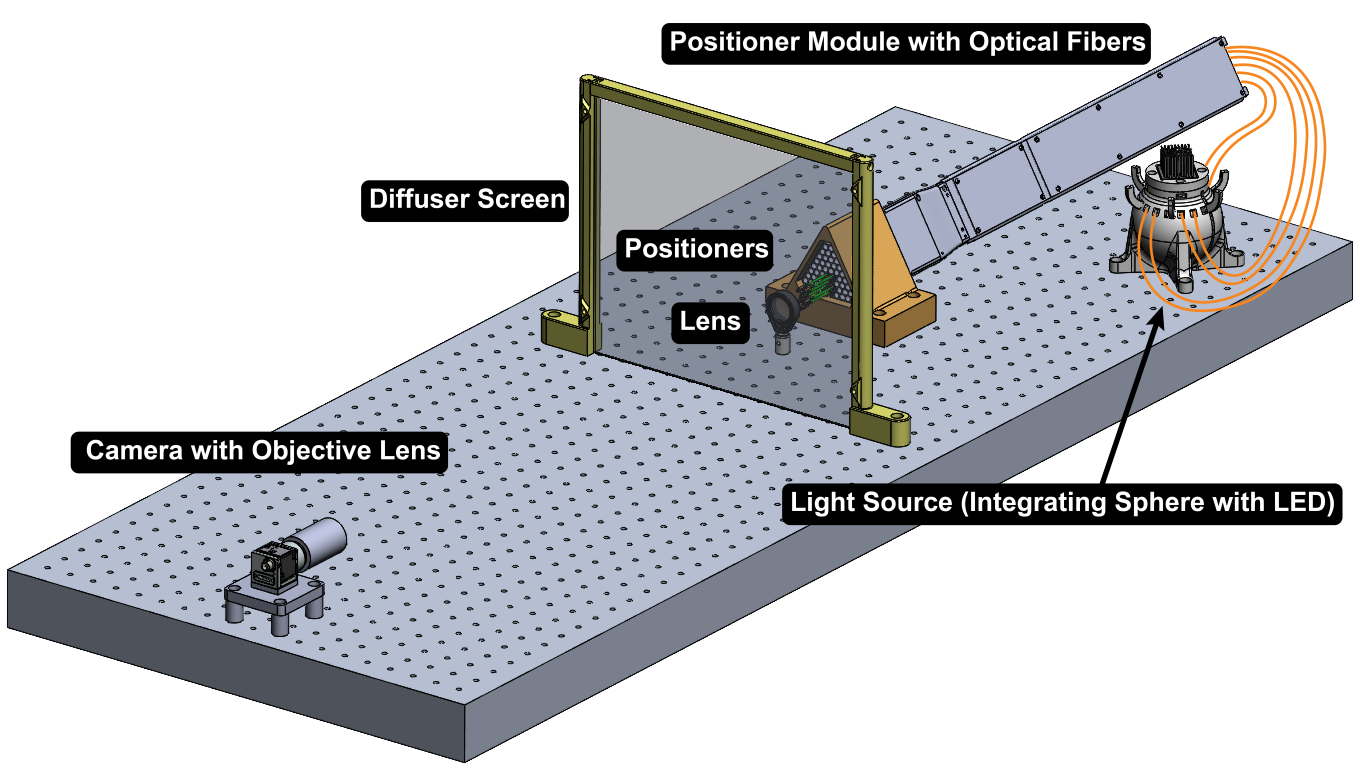}
    \caption{Rendered 3D model of the angular tilt test-bench using a short focal length lens (Transmission Setup).}
    \label{fig:shortFL_setup}
\end{figure}

The experimental test-bench utilized to measure the angular tilt stack (tilt between the optical fiber and the beta-arm, tilt between the beta-arm and the alpha-arm, tilt between the alpha-arm and the zero-reference) is shown in Figure \ref{fig:shortFL_setup}. The setup has been implemented based on the work presented by Kronig et al. (\cite{kronig_optical_2020, kronig_precision_2020}). It consists of a light source directly followed by a lens (bi-convex or plano-convex) of 100 mm focal length. The light then propagates through a 250x300 mm diffuser screen made of opal diffusing glass ($\#$43-042) which is placed at the focal length of the lens to ensure the disentanglement between the XY translation and tilt motions. The diffuser screen scatters the incoming light in many directions and results in a uniform illuminated spot. The use of the diffuser screen is highly necessary, as the diffuser acts as an angular scatterer which enables \textit{the full disentanglement of the light source's translation and tilt}. It is essential to differentiate between the two effects to be able to extract each of them separately. The illuminated spot projected on the screen is then captured by a 10 MP (acA3800-14um) Basler camera with a Fujinon objective lens (HF3520-12m) of a 35 mm focal length lens. The camera is placed about 650 mm away from the diffuser screen to ensure that the field of view covers the entire illuminated spot.

\subsubsection{Long Focal Length Test-bench (Reflection Setup)}
\label{sec5: long FL setup}

\begin{figure}[!b]
    \centering
    \includegraphics[width=0.7\linewidth]{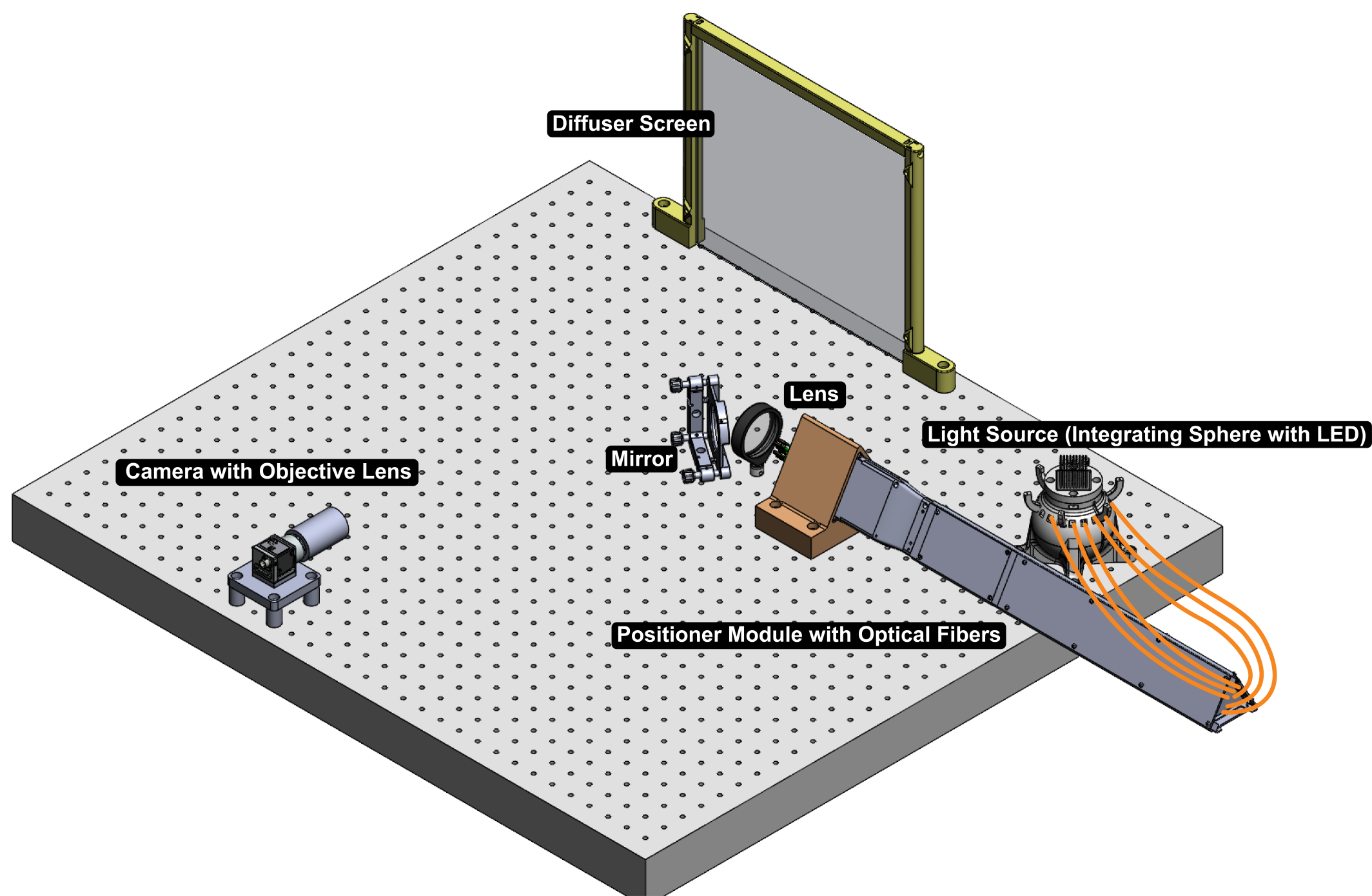}
    \caption{Rendered 3D model of the angular tilt test-bench using a long focal length lens (Reflection Setup).}
    \label{fig:longFL_setup}
\end{figure}

The setup presented earlier in Figure \ref{fig:shortFL_setup} works very well, but the disentanglement resolution of the system is limited, as it depends on the focal length of lens used. To be able to improve the system's resolution, the focal length chosen needs to be enlarged. That said, the experimental setup will automatically extend and will require a larger optical table. To be able to solve this problem, the original experimental test-bench by Kronig et al. (\cite{kronig_optical_2020, kronig_precision_2020}) has been modified (see Figure \ref{fig:longFL_setup}) to be able to fit on the optical table and still use multi-mode fibers in the positioners instead of single-mode fibers. The reason why it is important to utilize multi-mode fibers in the testing instead of single-mode ones is because ultimately the fibers which will be used in the surveys are multi-mode. So, to have a better understanding of the overall performance of the positioners, it is preferred to utilize similar fibers to those installed later when the positioner modules are mounted on the telescope.\\

The experimental setup (shown in Figure \ref{fig:longFL_setup}) includes the module with the optical fibers connected to the illumination source which is a powerful LED from Thorlabs (MINTF4). The optical fibers used are FBP polymicro fibers with 107 µm core diameter, 140 µm cladding diameter, and 170 µm polyimide coating. A plano-convex lens with focal length of 600 mm is directly placed after the positioners with the convex side facing the positioners. The current lens used covers all positioners. A flat mirror reflects the diverging light coming through the lens and projects it on a viewing screen larger than the opal diffuser used in the previous setup because the beam divergence is larger. In this modified system, we have the freedom to use any type of viewing screen which does not necessarily need to be transparent, as we use it in reflective mode. The viewing screen is placed at the focal length of the lens to ensure the disentanglement between the XY translation and tilt motions. A 20 MP Basler camera (acA5472-17um) having a larger field-of-view with a Fujinon objective lens (HF3520-12m) of a 35 mm focal length lens captures the illuminated spot projected on the screen. The camera is placed about 1200 mm away from the screen to ensure that the entire illuminated spot is acquired by the camera.

\subsection{Individual Axis Testing}

\begin{figure}[!t]
     \centering
     \captionsetup[subfigure]{justification=centering}
     \begin{subfigure}[b]{0.4\textwidth}
         \centering
         \includegraphics[width=\textwidth]{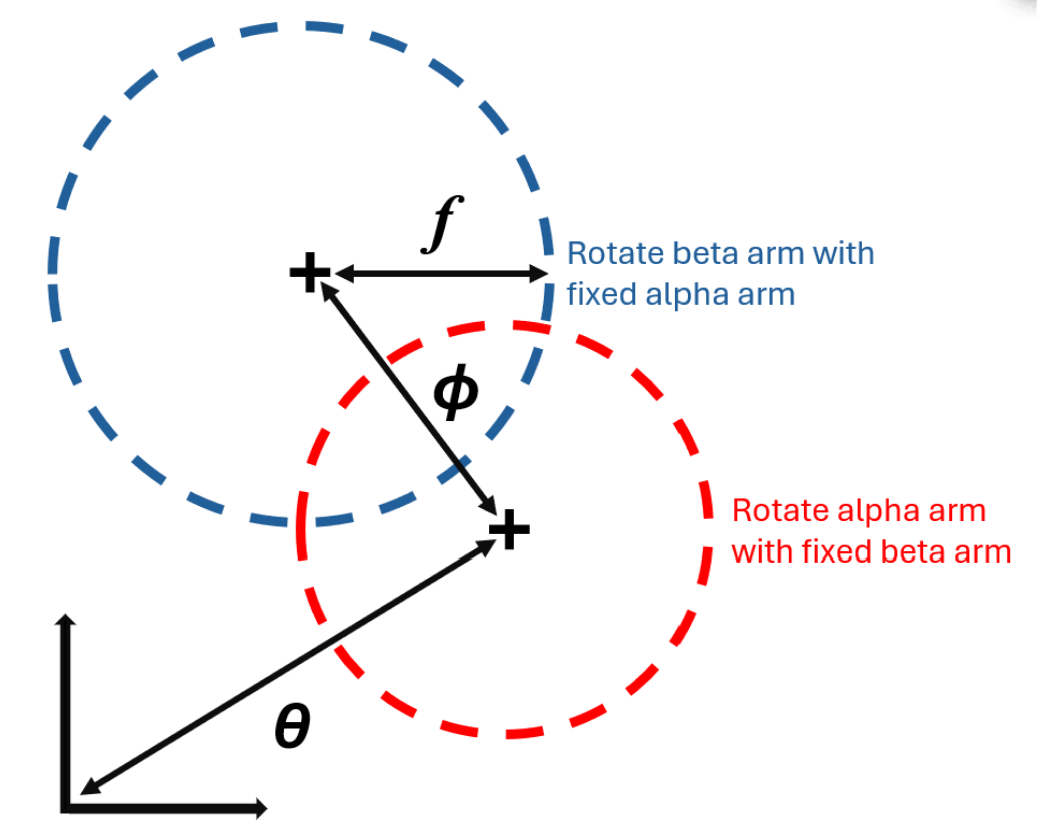}
         \caption{}
         \label{fig:tilt_circles}
     \end{subfigure}%
     \hspace{0.5cm}
     \begin{subfigure}[b]{0.35\textwidth}
         \centering
         \includegraphics[width=\textwidth]{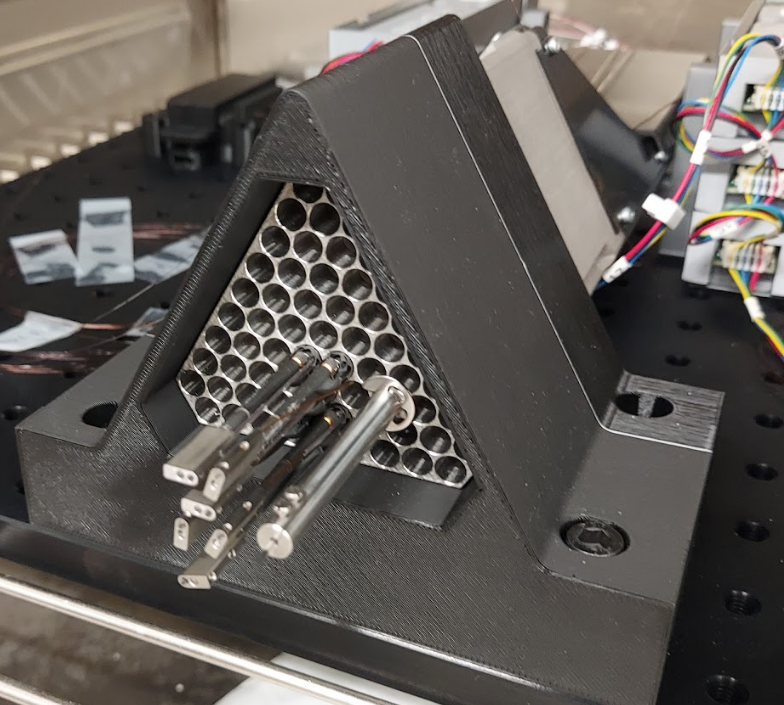}
         \caption{}
         \label{fig:calibration_cylinder}
     \end{subfigure}%
        \caption{(a) Diagram illustrating a breakdown of the tilt components and their orientations; (b) Calibration cylinder used to calculate the positioner zero-reference.}
\end{figure}

\begin{itemize}
    \item \textbf{Tilt between $FI_A\#$ and Ferrule Axis}\\
    Before installing the optical fibers on the robotic positioners, they were installed on a perfectly-round calibration cylinder, shown in Figure \ref{fig:calibration_cylinder}), and rotated to be able to measure the tilt of the fiber inside the ferrule. 

    \item \textbf{Tilt between Ferrule Axis and $Beta_A\#$ (denoted as \textit{f} in Figure \ref{fig:tilt_circles})}\\
    The tilt between the ferrule axis and the beta arm axis is measured by rotating the beta arm at several angles, fitting a circle, and measuring the radius of the circle.

    \item \textbf{Tilt between $Beta_A\#$ and $Alpha_A\#$ (denoted as $\phi$ in Figure \ref{fig:tilt_circles}))}\\
    The tilt between the beta arm axis and the alpha arm axis is measured by rotating the alpha arm at several angles while the beta arm is stationary, fitting a circle, and measuring its centroid. Then, rotating the beta arm at several angles while the alpha arm is stationary, fitting a circle, and measuring its centroid. Finally, calculating the angle difference between the centroids.

    \item \textbf{Tilt between $Alpha_A\#$ and $Pos_A\#$ (denoted as $\theta$ in Figure \ref{fig:tilt_circles})}\\
    The tilt between the alpha arm axis and the positioner axis is measured by having a calibration cylinder that is inserted in one of the empty module holes (as shown in Figure \ref{fig:calibration_cylinder})and acting as a non-tilted zero-reference. By rotating this calibration cylinder at different angles, measuring the centroid of the captured illuminated spot, and fitting a circle, the zero-reference is estimated as the center of the fitted circle.
    
\end{itemize}

\subsection{Image Post-processing}

\begin{figure}[!t]
     \centering
     \captionsetup[subfigure]{justification=centering}
     \begin{subfigure}{0.3\textwidth}
         \centering
         \includegraphics[width=\textwidth, height=4cm,keepaspectratio]{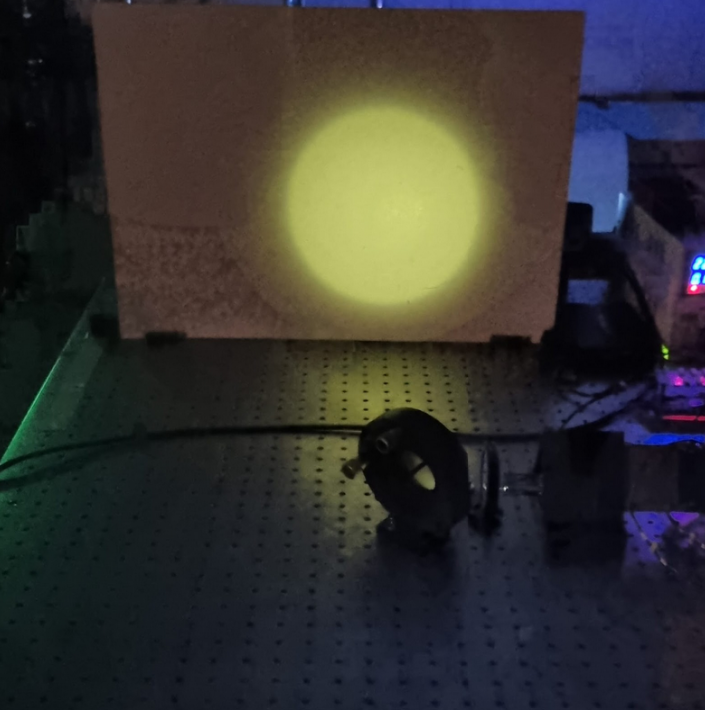}
         \caption{}
         \label{fig: blob on screen}
     \end{subfigure}%
     \hfill
     \begin{subfigure}{0.3\textwidth}
         \centering
         \includegraphics[width=\textwidth, height=4cm,keepaspectratio]{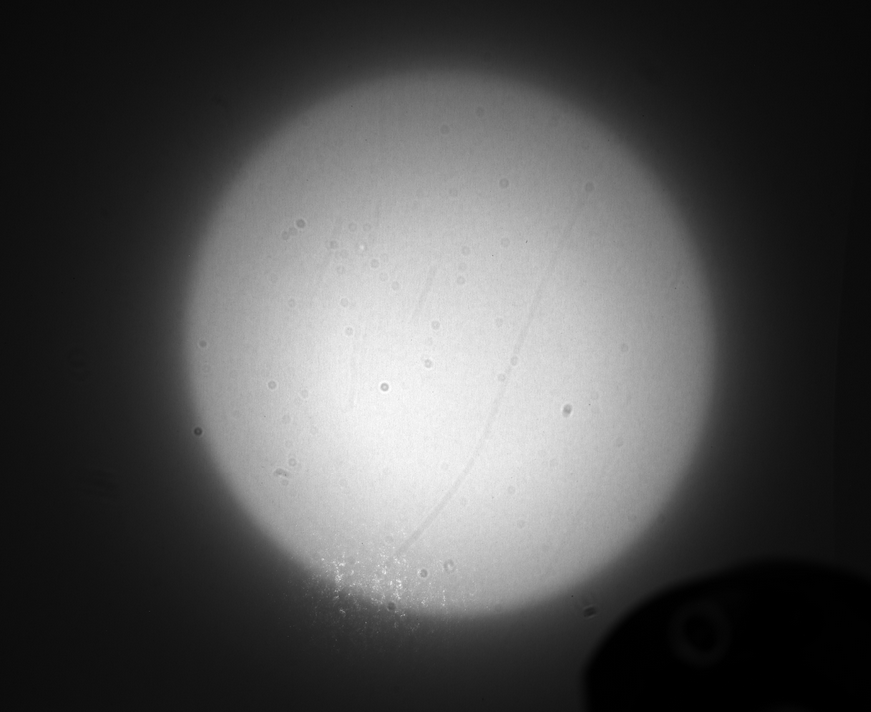}
         \caption{}
         \label{fig: blob on camera}
     \end{subfigure}%
          \hfill
     \begin{subfigure}[b]{0.3\textwidth}
         \centering
         \includegraphics[width=\textwidth, height=4cm,keepaspectratio]{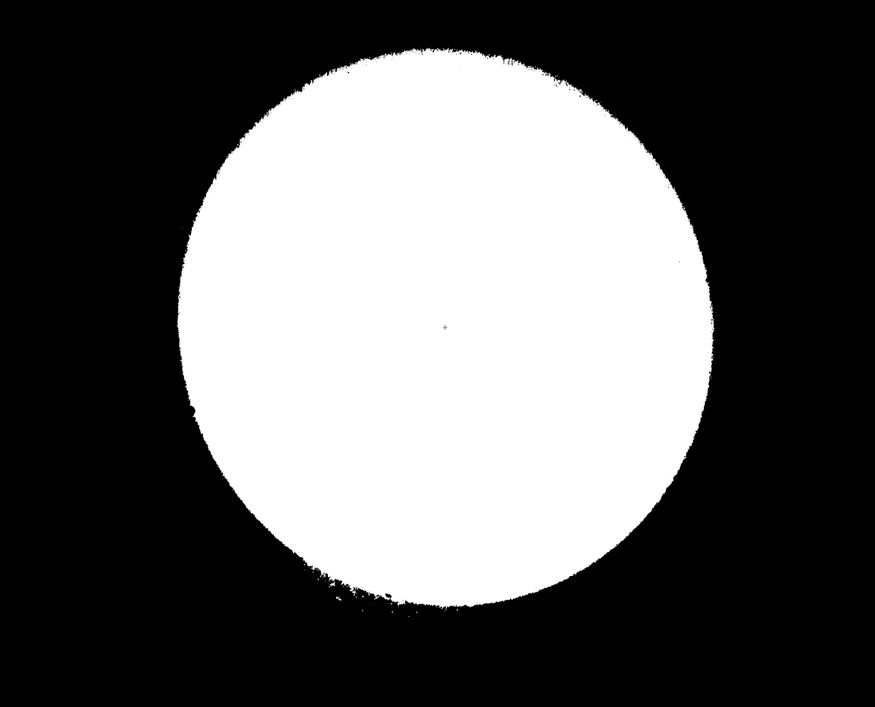}
         \caption{}
         \label{fig: blob thresholded}
     \end{subfigure}
        \caption{The illuminated spot at different stages; (a) on screen, (b) on camera, and (c) after post-processing.}
        \label{fig:blob}
\end{figure}

The illuminated spot projected on the diffuser screen is shown in Figure \ref{fig: blob on screen}, its image captured by the camera is shown in Figure \ref{fig: blob on camera}, and after post-processing, the illuminated spot is shown in Figure 8 \ref{fig: blob thresholded}.

The raw image of the illuminated spots undergoes post-processing which includes Gaussian blurring and Otsu’s thresholding to be able to calculate the centroid of the illuminated spot. 
When the alpha and beta arms are rotated, the centroids of the illuminated spots create a circle which is fitted using a least squares fit. The fit finds the best-fit circle center and radius such that the points approximately lie on the circle.

\subsection{Angular Tilt Test Results}
    \begin{figure}[!b]
    \centering
    \includegraphics[width=0.7\linewidth]{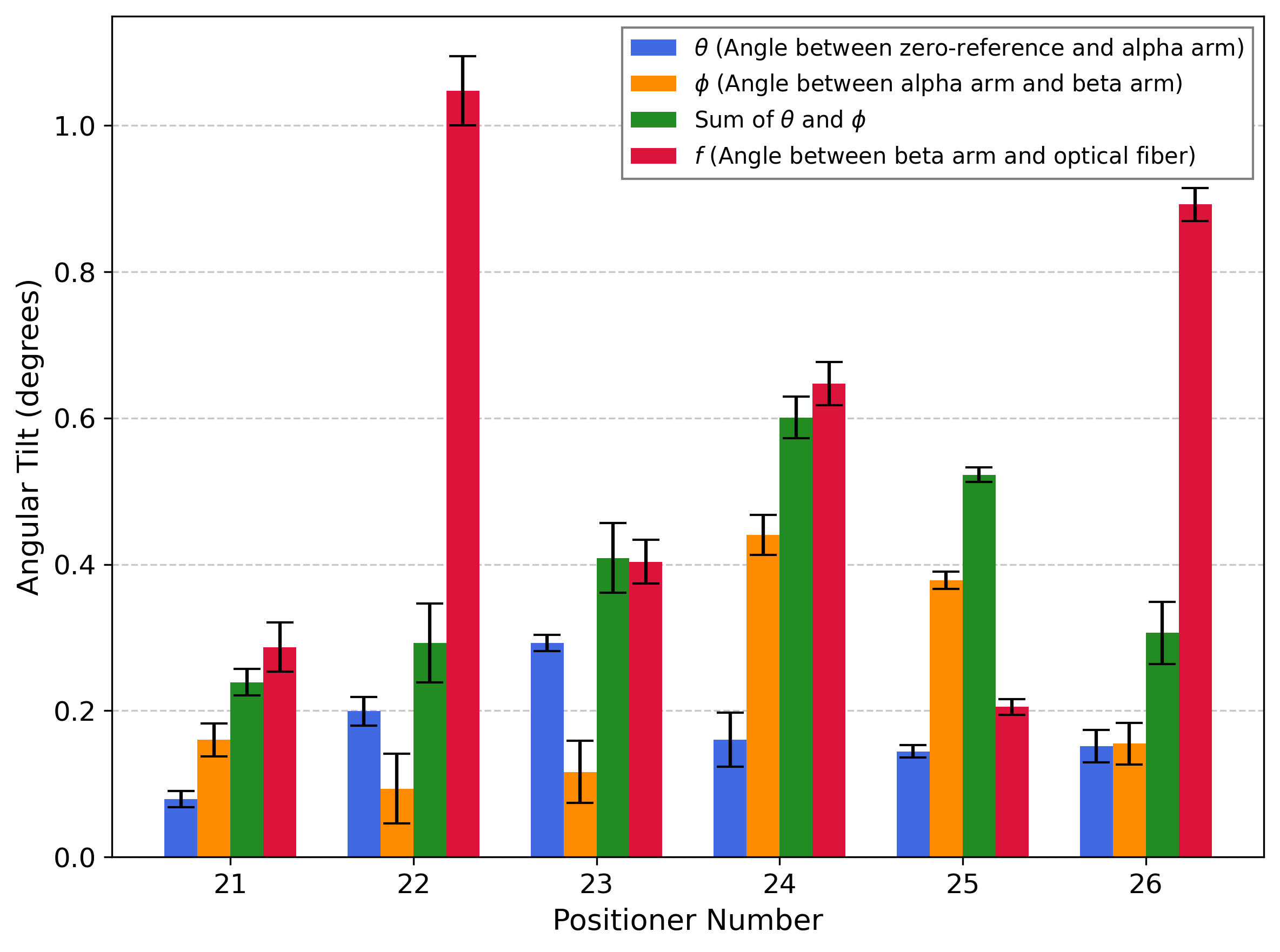}
    \caption{Graph showing the angular tilt of each of the positioner axes for the 6 robotic positioners of MPS Prototype 1.}
    \label{fig:tilt_results}
\end{figure}
The angular tilt test results are presented in Figure \ref{fig:tilt_results}. It is a bar plot that shows the different tilt contributions for each of the 6 positioners tested. The blue bar shows theta representing the angle between the zero-reference of the positioner and the alpha arm axis; the orange bar shows phi representing the angle between the alpha arm axis and the beta arm axis; the green bar represents the summation of $\theta$ and $\phi$; and the red bar shows $f$ which represents the angle between the beta arm axis and the optical fiber axis. The calculation of the results was done by constructing alpha and beta circles each containing five data points to fit the circles and the program was repeated three times. 

The summation is intended to show the tilt result of both $\theta$ and $\phi$ while taking into account the worst case scenario of the tilt contributions. This clarifies whether the tilt contributions will be summed up or will compensate each other. As can be seen in Figure \ref{fig:tilt_results}, the summation of $\theta$ and $\phi$ of all positioners is very close to 0.5° of tilt which is a very positive indication on the performance of the positioners in that aspect. Certainly, requirements vary from project to project, but it is worth highlighting that the results are definitely promising considering that this is the first round of prototyping and further improvements to the arms are foreseen.

Another important parameter to consider in the overall tilt is the tilt between the beta arm and the optical fiber. As mentioned earlier in Section \ref{sec:angular_tilt_axes}, the contribution of this tilt is a combination of the tilt of the optical fiber inside the ferrule and the ferrule inside the beta arm. Therefore, it is highly critical to make sure that the process of the optical fiber insertion in the ferrule is very meticulously performed. The optical fibers used in our experiment were available fibers and not intentionally prepared, but it is recommended to use better fibers for future prototypes.

\section{Experimental Testing on Two Orbray Prototypes}
In collaboration with the team of Lawrence Berkeley National Lab (LBNL), two Orbray prototypes with 6 positioners each (2 Trillium) were tested. Orbray Prototype 1 was tested at EPFL and Orbray Prototype 2 was tested at LBNL. 

\subsection{Basic Trillium Mechanics}
Each Trillium unit has three positioners that each have two axes, alpha and beta. The alpha and beta axes are coupled via a transfer gear (see Figure \ref{fig:trillium_robots} and Silber et al (2022) (\cite{silber_25000_2022}) for more information). For moves with the beta arm, we define a move as changing the beta angle only. For moves with the alpha arm, we define moves as changing the alpha angle while keeping the beta angle constant. Because the two arms are coupled, when the alpha angle is changing the beta angle will also change. In order to keep the beta angle constant, the beta motor must be commanded to move in equal magnitude but opposite direction to the alpha motor. Since the motors used in the Trillium are the same as the DESI fiber positioners, the team at LBNL uses DESI electronics and operates the motors at either of the two DESI speeds: 150 RPM or 9900 RPM (\cite{silber_robotic_2023}). The team at EPFL uses the same electronics used for SDSS-V positioners. 

\subsection{Test Setup}
\begin{figure}[!t]
     \centering
     \captionsetup[subfigure]{justification=centering}
     \begin{subfigure}[b]{0.565\textwidth}
         \centering
         \includegraphics[width=\linewidth]{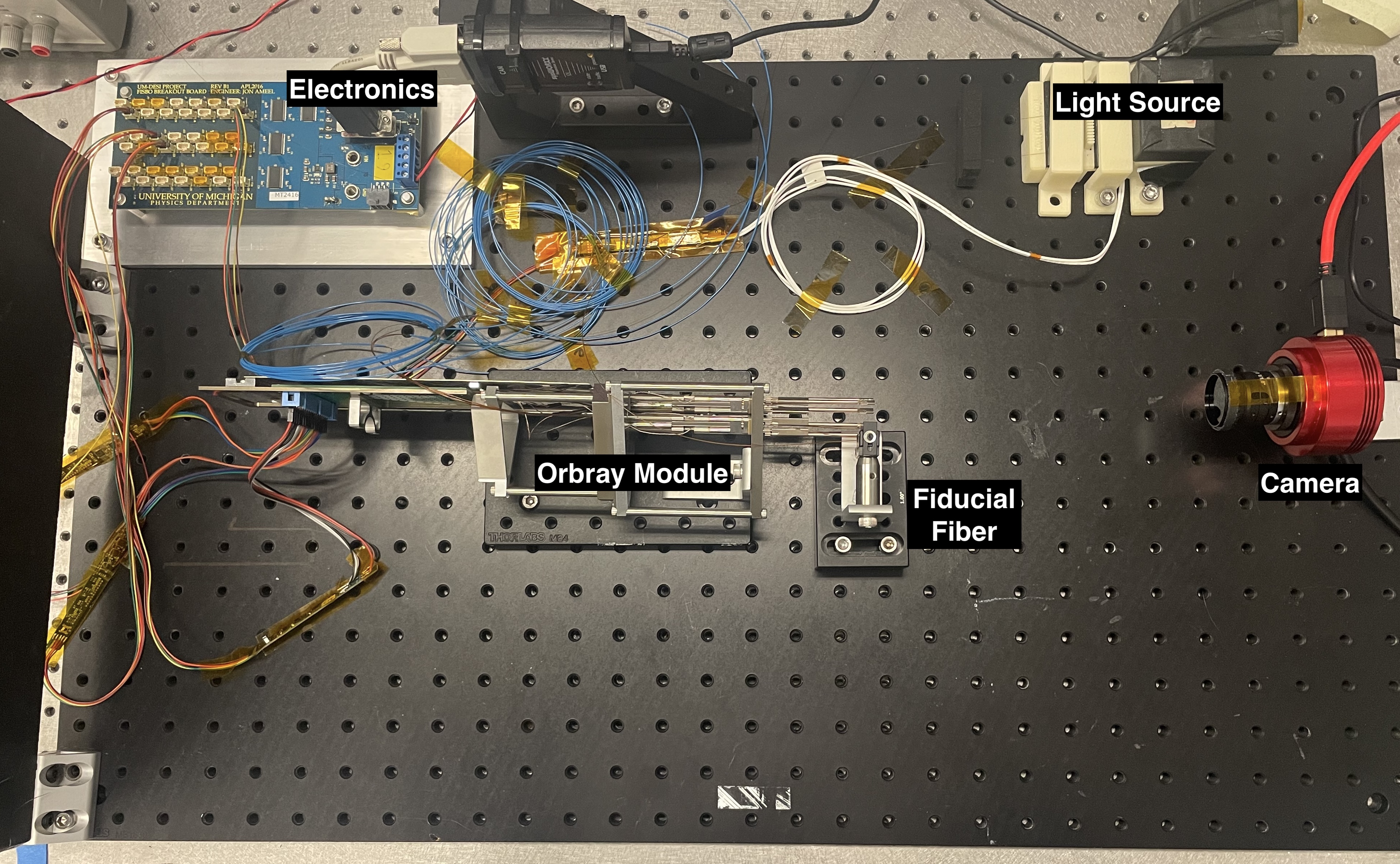}
    \caption{}
    \label{fig:lbnl-setup}
     \end{subfigure}%
     \hspace{0.5cm}
     \begin{subfigure}[b]{0.29\textwidth}
         \centering
         \includegraphics[width=\linewidth]{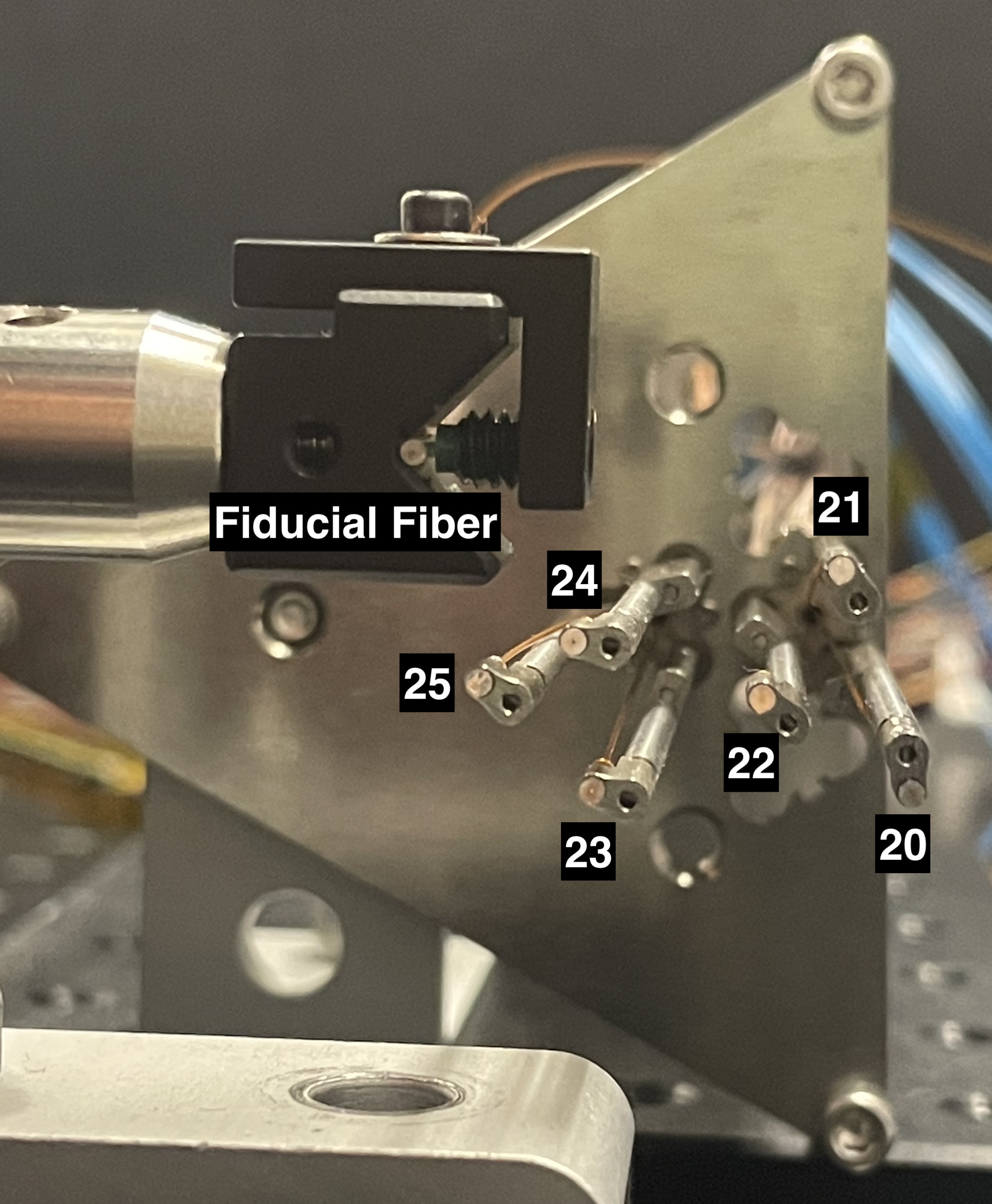}
    \caption{}
    \label{fig:lbnl-raft-setup}
     \end{subfigure}%
        \caption{(a) Experimental test-bench by LBNL; (b) Positioner IDs for the 6 positioners tested on the LBNL Orbray module.}
\end{figure}

The experimental test setup used by EPFL is the same as that presented in Figure \ref{fig:xy_testbench}. Similarly, LBNL mounted the test-bench in their lab. The setup shown in Figure \ref{fig:lbnl-setup} measures the X,Y positions of fibers via a 6.4 MP ZWO camera (ASI178MM). Attached to the camera is a 470 nm bandpass filter. 
The module is mounted horizontally to the optical bench and a fiducial fiber is added to remove instability in the setup. The fiducial and positioner fibers are back-lit using an integrating sphere that is connected to a 470 nm LED. Positioners are individually commanded to move to different locations, an image is taken before and after each move, and centroids are measured to determine the X,Y positions of the positioner and fiducial fibers. This setup is used for two main tests discussed in sections \ref{subsec:lbl_arcs} 
and \ref{subsec:lbl_backlash}, although it could be adapted for other tests as well.

\subsection{XY Repeatability Measurement}
For the positioning repeatability measurement, the data is acquired by moving the alpha arms from 0 to 360 degrees with a step of 30 degrees, and the beta arms from 0 to 180 degrees in steps of 30 degrees multiple times. The centroids of the light illuminated spots are found using 2D Gaussian fitting, and the root mean square of the repeatability is calculated from 10 iterations of reaching each XY position. The data presented in Figure \ref{fig: x-y repeatability} shows the mean of the root mean square values of the program repeated 10 times. The figure presents the repeatability data of the two Orbray prototypes. Prototype 1 has been tested by EPFL and Prototype 2 has been tested by LBNL. It should be noted that the disparity in the results between the two prototypes may be because the EPFL prototype has experienced several runs whereas the LBNL prototype has only undergone a few runs. With more runs, it was noticed that the repeatability of the positioners improves over time, which can be explained by the mechanics being looser after several runs. The EPFL prototype has been tested for XY positioning repeatability, and the rest of the testing is still to be conducted and is, therefore, out of the scope of this paper.

\begin{figure}[!t]
     \centering
     \captionsetup[subfigure]{justification=centering}
     \begin{subfigure}[b]{0.5\textwidth}
         \centering
         \includegraphics[width=\textwidth]{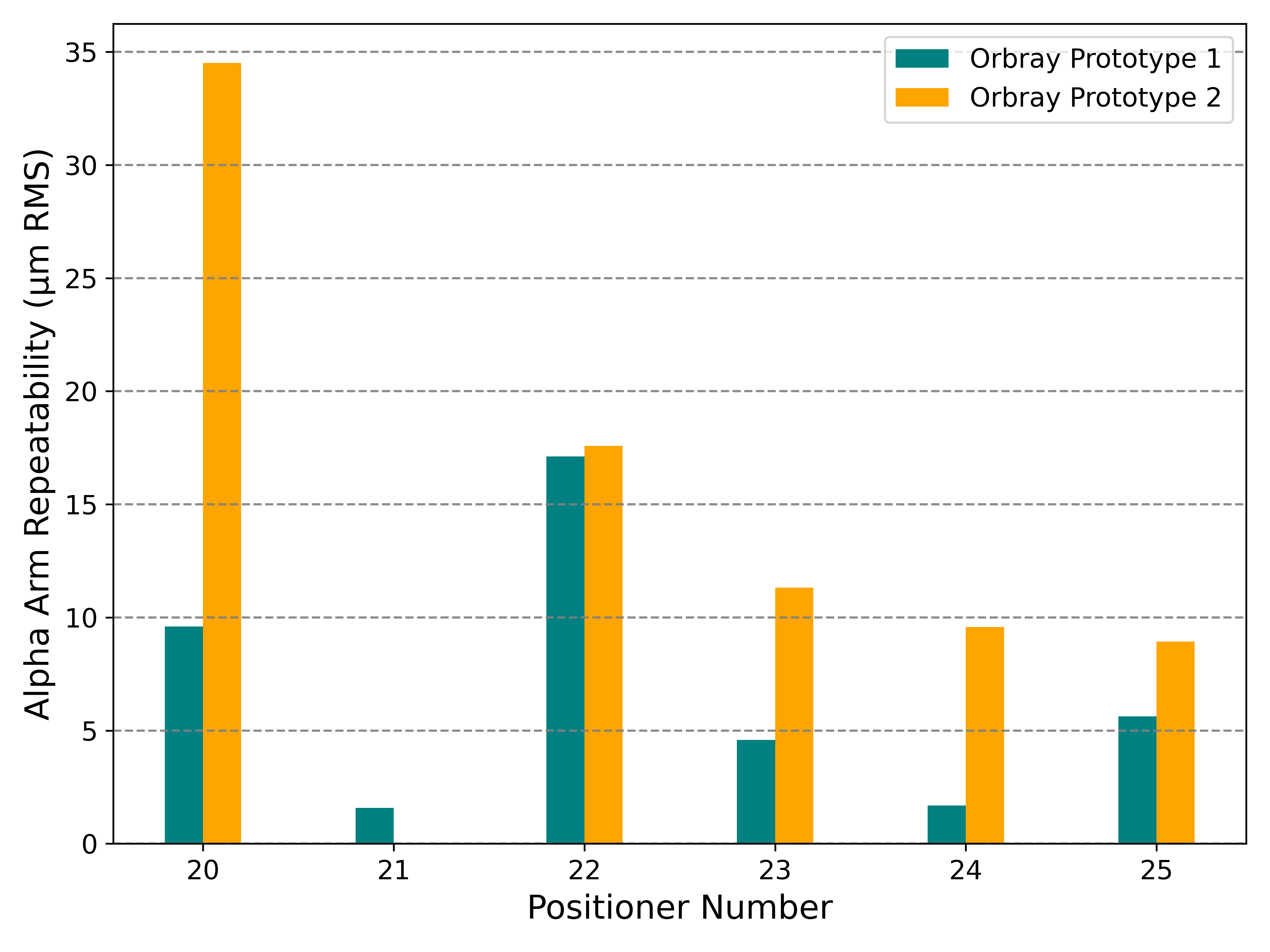}
         \caption{Alpha Arm}
         \label{fig: alpha_arm_repeatability_Orbray}
     \end{subfigure}%
     \hfill
     \begin{subfigure}[b]{0.5\textwidth}
         \centering
         \includegraphics[width=\textwidth]{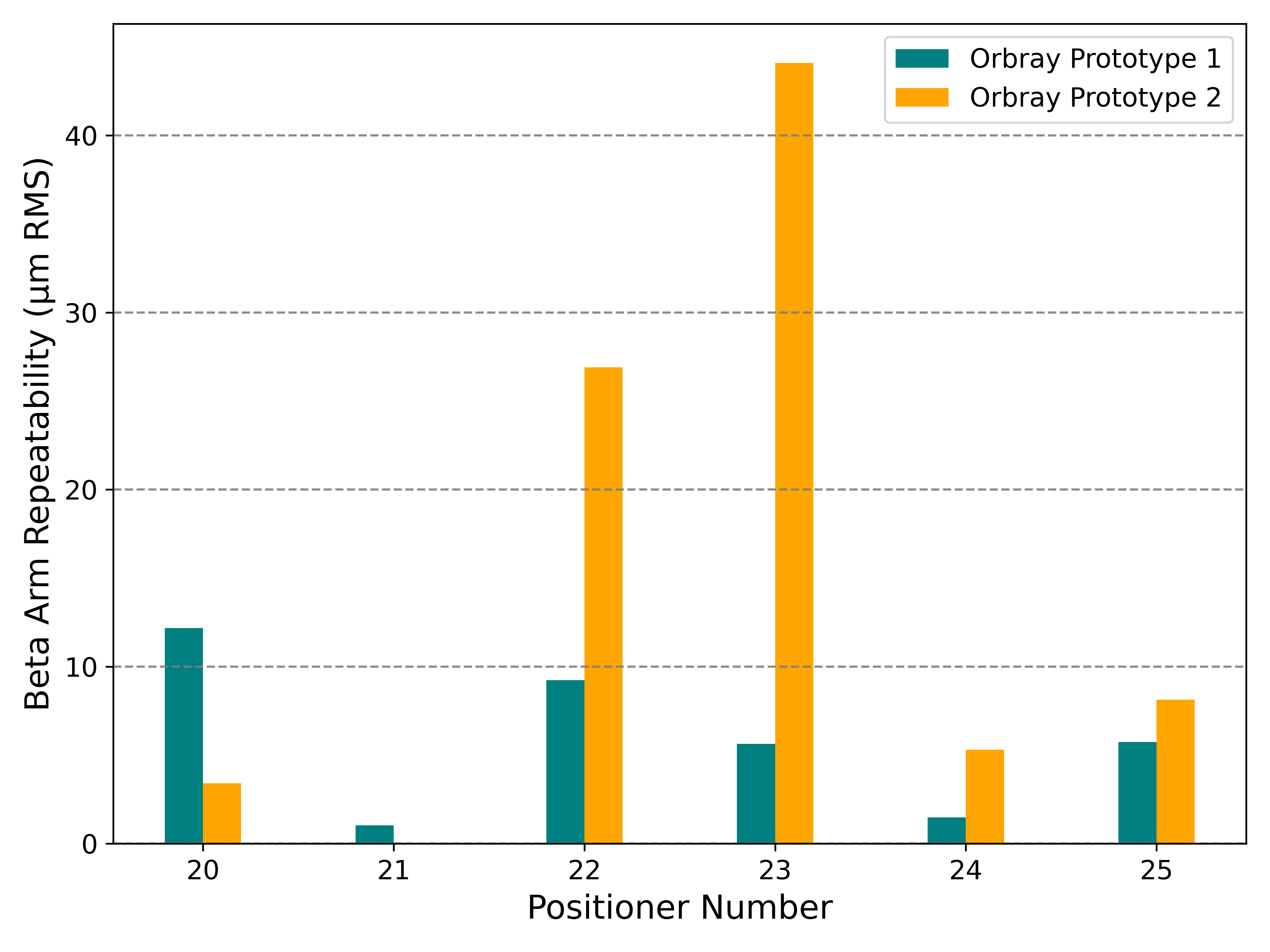}
         \caption{Beta Arm}
         \label{fig: beta_arm_repeatability_Orbray}
     \end{subfigure}%
        \caption{Graphs showing the \textbf{XY positioning performance repeatability} for two Orbray prototypes. The bar plots show the root-mean-square values versus the positioner number; (a) results for the alpha arm, and (b) results for the beta arm.}
        \label{fig: x-y repeatability Orbray}
\end{figure}

\subsection{High Resolution Arc Tests}
\label{subsec:lbl_arcs}

\begin{figure}[!t]
     \centering
     \captionsetup[subfigure]{justification=centering}
     \begin{subfigure}[b]{0.388\textwidth}
         \centering
         \includegraphics[width=\textwidth]{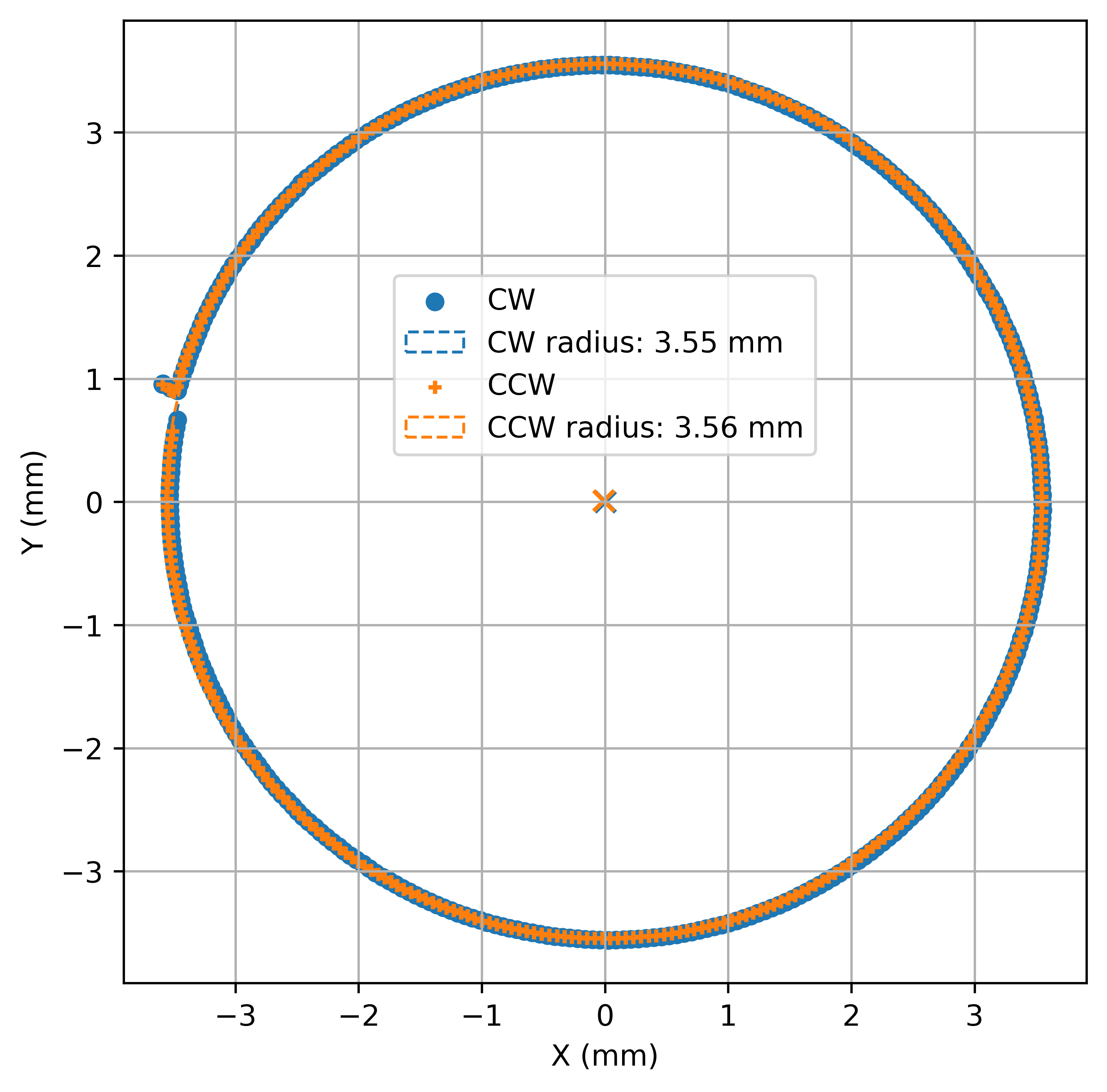}
         \caption{Alpha Circle}
         \label{fig: alpha_nl_circle2}
     \end{subfigure}%
     \hspace{0.5cm}
     \begin{subfigure}[b]{0.4\textwidth}
         \centering
         \includegraphics[width=\textwidth]{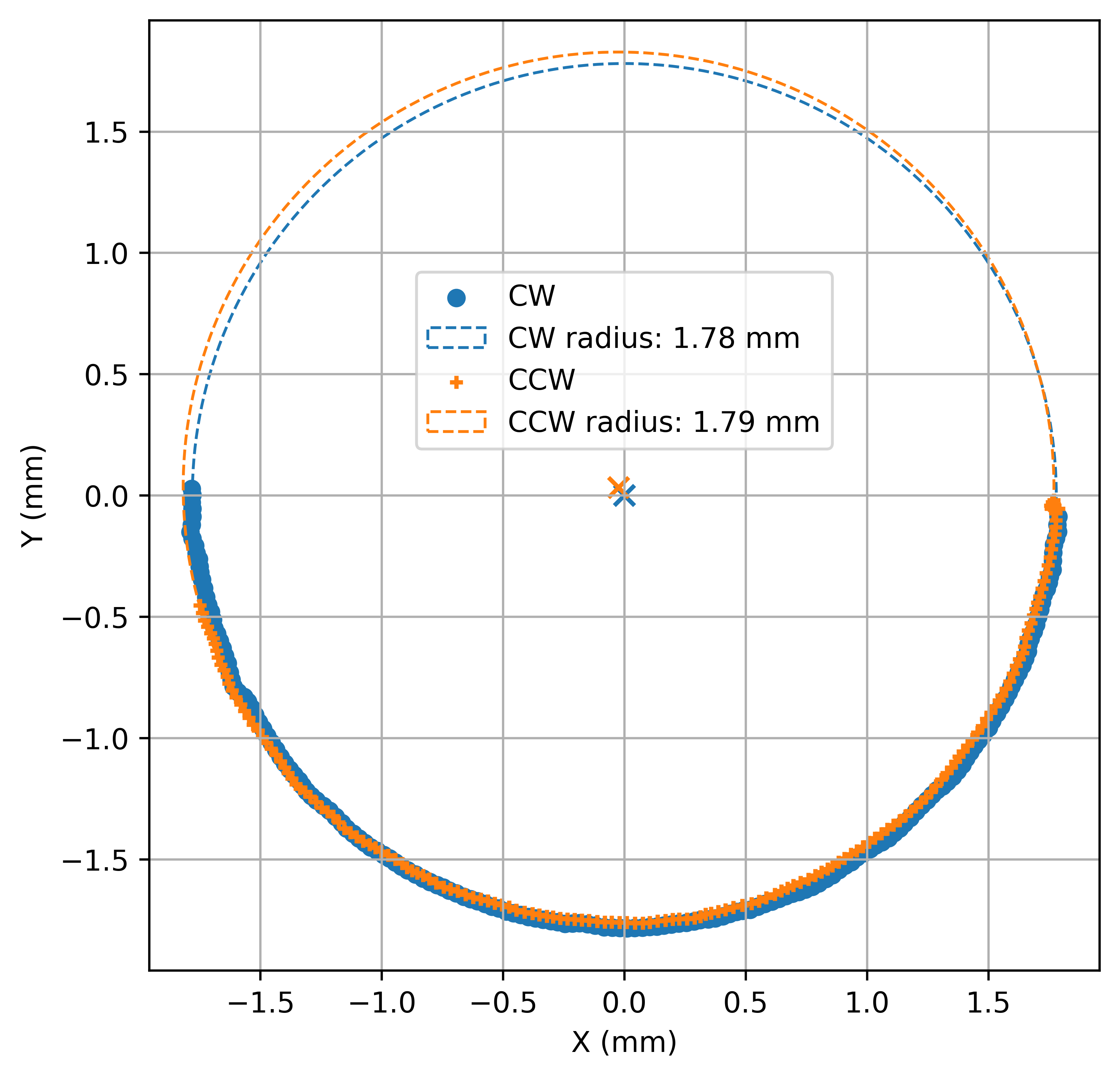}
         \caption{Beta Arc}
         \label{fig: beta_nl_arc2}
     \end{subfigure}%
        \caption{Normalized centroid locations for positioner 21 during the high resolution arc tests for the alpha and beta axes. One degree moves are commanded spanning the full range of each axis in the clockwise direction (CW; blue points) and then in the counter-clockwise direction (CCW; orange points).}
        \label{fig:lbnl-posid21-arcs}
\end{figure}

The team at LBNL conducted further testing on the Orbray Prototype 2. The high resolution arc tests performed at LBNL were designed to see positioner movement over the range of each axis. The nominal ranges of alpha and beta are 360 and 180 degrees, respectively (\cite{silber_25000_2022}). We perform one degree moves with a motor speed of 150 RPM that span these ranges in both the clockwise and counter-clockwise directions. An example of the centroid positions is shown in Figure \ref{fig:lbnl-posid21-arcs} for positioner 21. We fit a circle to moves in each direction to measure the center. 
The centers are then used to calculate the angle between each move and is saved to calculate the angles during the backlash tests discussed in the next section. 

Nominally, the center is a fixed location, however we see some deviation in the measured centers in each direction. This is seen in the beta arc of Figure \ref{fig:lbnl-posid21-arcs} where the centroid positions are different for each direction. 
The length of the beta arm is fixed, but the center of the beta arm will depend on the location of the alpha arm - it will be fixed if alpha is not moving - and it's possible that, though it was not being commanded, the alpha arm moved slightly during the beta arc test for this positioner. 

The positioning results for the 6 positioners are shown in Table \ref{table:lbl_results} in the "Arc Residuals" column. The residual is calculated as the target angle minus the measured angle. The RMS for each positioner in each axis is shown, and is well within 0.3 degrees for all positioners. The performance in the alpha axis is generally better than the performance in the beta axis, likely due to the coupled nature of the Trillium design. 

\begin{table}[!t]
    \centering
    \small
    \renewcommand{\arraystretch}{1.3} 
    \begin{tabular}{|c|*8{w{c}{1.5cm}|}} \hline
        \multirow{3}{*}{\shortstack{Positioner \\ ID}} & 
        \multicolumn{2}{c|}{\parbox[t]{3.2cm}{\centering Backlash [deg]\\ (Orbray Prototype 1)}} & 
        \multicolumn{2}{c|}{\parbox[t]{3.2cm}{\centering Backlash [deg]\\ (Orbray Prototype 2)}} & 
        \multicolumn{2}{c|}{\parbox[t]{3.2cm}{\centering Arc Residual [$\mu m$] \\ (Orbray Prototype 1)}} & 
        \multicolumn{2}{c|}{\parbox[t]{3.2cm}{\centering Arc Residual [$\mu m$] \\ (Orbray Prototype 2)}} \\ \cline{2-9}
         & Alpha & Beta & Alpha & Beta & Alpha & Beta & Alpha & Beta \\ \hline
         20 & 0.11 & 16.67 & 10.22 & 13.20 & 7.59 & 2.09 & 5.03 & 8.17 \\ \hline
         21 & 1.55 & 14.61 & 10.79 & 12.70 & 5.63 & 6.60 & 5.65 & 5.97 \\ \hline 
         22 & 0.07 & 7.66 & 4.53  & 10.60 & 5.54 & 7.97 & 13.19 & 7.85 \\ \hline 
         23 & - & - & 5.92  & 11.40 & -    & -    & 6.28 & 7.54 \\ \hline 
         24 & - & - & 6.07  & 12.42 & 2.47 & 4.18 & 5.65 & 5.97 \\ \hline 
         25 & - & - & 6.02  & 10.24 & 2.54 & 13.99 & 8.17 & 7.23 \\ \hline
    \end{tabular}
    \vspace{5pt}
    \caption{RMS of the measured backlash (in degrees) for each positioner in the alpha and beta axes. The backlash is calculated over 200 20 degree moves, alternating between clockwise and counter-clockwise directions. The arc residuals are calculated as the target angle minus the measured angle over 1 degree moves that span the range of each axis. The results shown in the table are for two Orbray prototypes where the Prototype 1 has been tested at EPFL and Prototype 2 has been tested by LBNL. Note that only one trillium (positioners 20, 21, 22) in the Orbray Prototype 1 has been tested for backlash. Also, note that Positioner 23 of Prototype 1 exhibited a significant bend of the arms leading to unrealistic arc residual results and thus were neglected in the comparison.}
    \label{table:lbl_results}
\end{table}


\subsection{Backlash Tests}
\label{subsec:lbl_backlash}

The backlash tests performed at LBNL are similar to those performed at EPFL (described in section \ref{subsec:performance_metrics}). In the tests at LBNL, we commanded each positioner to move 20 degrees, alternating between clockwise and counter-clockwise directions, 100 times in each direction at the 9900 RPM motor speed. We use the center points measured in section \ref{subsec:lbl_arcs} to calculate the angle between each move, called the measured angle. The measured backlash is then calculated as the target angle (20 degrees) minus the measured angle, assuming that the amount each positioner does not move is solely due to the gear backlash. As mentioned in section \ref{subsec:performance_metrics}, lower backlash values allow more precise prediction of the path of motion. This is important for spectroscopic surveys as it affects how densely astronomical targets can be selected. 

The results of the backlash tests for the Orbray 6-positioner prototypes are shown in Table \ref{table:lbl_results} in the ``Backlash'' column where the RMS for each positioner is calculated. 
As can be seen in the table the values for the alpha arms vary quite significantly when comparing Prototype 1 and Prototype 2. It should be noted that Prototype 1 was measured at EPFL and Prototype 2 was measured at LBNL. The results for the beta arm of both prototypes are in the same order of magnitude. The disparity in the alpha arm between the two prototypes could either be a difference in the mechanics or a difference in the estimation given the complexity of the coupled mechanics of the trillium.


\subsection{Tip-Tilt Measurement}

\begin{figure}[!t]
    \centering
    \includegraphics[width=0.5\linewidth]{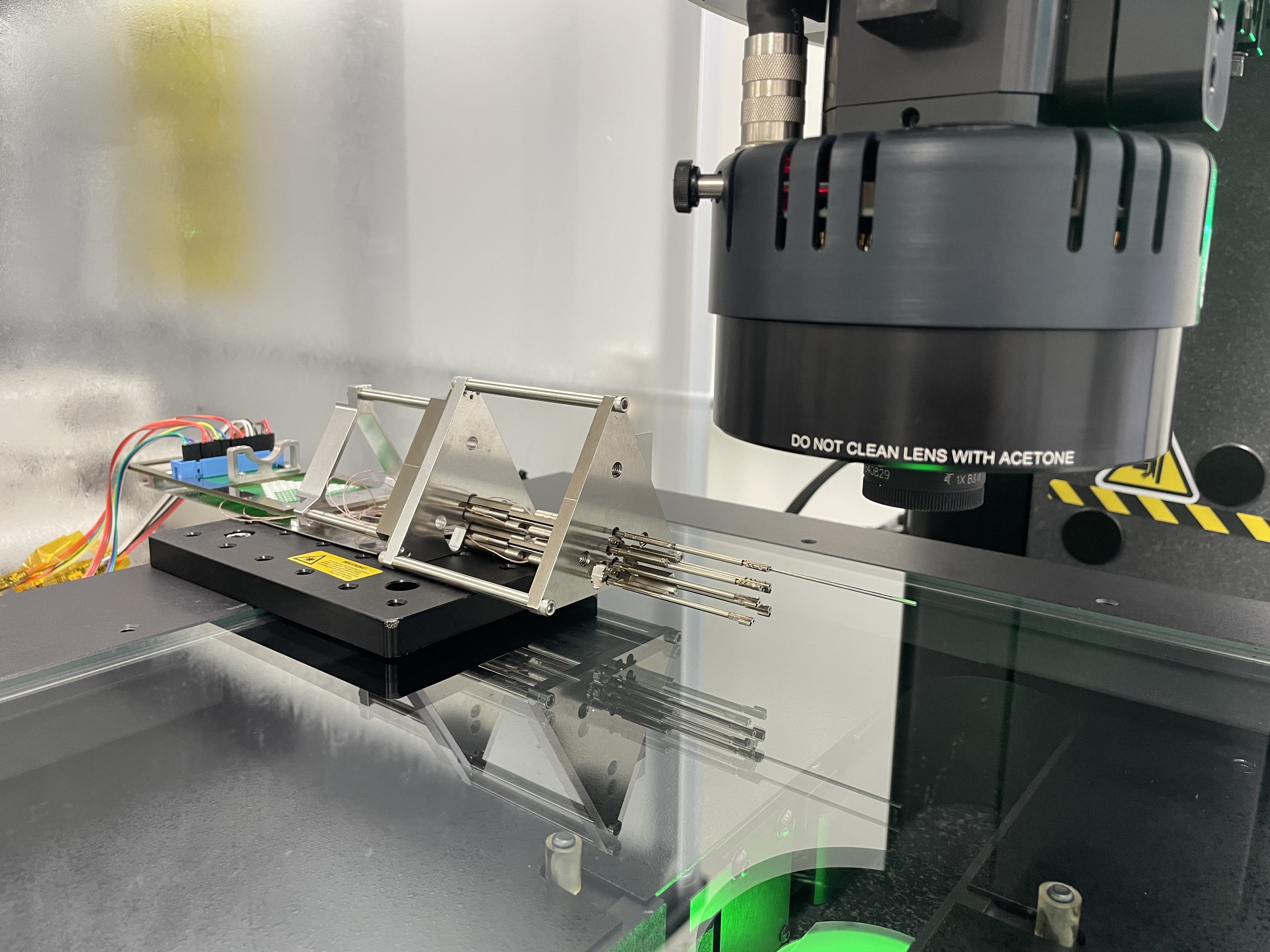}
    \caption{Angular tilt setup on the LBNL CMM.}
    \label{fig:lbnl-cmm}
\end{figure}

\begin{table}[!b]
    \centering
    \begin{tabular}{|c|c|c|c|c|c|} \hline
         Pos ID & $\theta$ & $\phi$ & f & focus$_\text{AB}$ & focus$_\text{CD}$ \\ \hline
         20 & -0.29 & -0.34 & -0.22 & -0.2 & -11.5 \\
         21 & -0.26 & -0.20 & -0.39 & -6.1 & 3.3 \\
         22 & -0.33 & -0.40 & -0.34 & 1.8 & 0.4 \\
         23 & -0.31 & -0.50 & -0.14 & -190 & -40 \\
         24 & -0.23 & 0.05 & -0.78 & -9.4 & -70 \\
         25 & -0.11 & -0.28 & -0.20 & -30 & -10 \\ \hline
         RMS & 0.26 & 0.33 & 0.40 & 79 & 33 \\ \hline
         
    \end{tabular}
    \caption{Positioner axis measurements (degrees), and focus ($\mu$m) measurements from the LBNL optical CMM measurements. }
    \label{tab:lbl-tip-tilt}
\end{table}

We measure the tip and tilt angles of the LBNL Orbray module (Prototype 2) using an optical coordinate measuring machine (CMM) at LBNL. There are three angles that we measure: the angle between ferrule and the beta arm ($f$), the angle between the beta and alpha arms ($\phi$), and the angle between the alpha arm and the positioner axis ($\theta$), which are all described in table \ref{tab:tilt_axes} and shown in Figure \ref{fig:tilt_circles}. For our measurements, we remove the fiber and instead use a long ferrule rod. On the CMM the measurements are made with respect to the module axis, and we assume that the positioners are aligned to the module.

We can only take measurements from the top-down view on the CMM, however we are still able to measure the angles $\theta$, $\phi$, and $f$. We are also able to calculate the change in focus by measuring the location of the tip of the ferrule rod. An example of the setup is shown in Figure \ref{fig:lbnl-cmm}. We take four measurements for each positioner in the following order:
\begin{enumerate}[label=\Alph*.,noitemsep,topsep=1pt,leftmargin=3\parindent]
    \item Alpha at hardstop, beta fully extended
    \item Move alpha 180 degrees, keeping beta fully extended
    \item Alpha at hardstop, beta fully extended
    \item Move beta 180 degrees.
\end{enumerate}
We then rotate the module 90 degrees and repeat these measurements for each positioner.


From these measurements, we are able to calculate the angles (shown in Figure \ref{fig:tilt_circles}) as well as getting a measurement of the focus between points A and B and points C and D listed previously. The focus measurement is just the difference in the position of the end of the ferrule rod at each point. The results are shown in table \ref{tab:lbl-tip-tilt}. Across the measured range of positions, the change in focus ranged from 0.16 - 190 $\mu$m and the change in angular alignment of individual axes ranged from 0.05 - 0.78 degrees.


         

\section{Discussion and Conclusion}
The results presented in this work aim to highlight the performance of the first modules of 6.2-mm-pitched robotic optical fiber positioners. The prototypes of different manufacturers (MPS and Orbray) have been tested and evaluated in a comparative manner. Each project has its own requirements, and some of the general specifications for such projects are mentioned in one of the very first references presenting the concept of Stage-V telescopes (\cite{silber_25000_2022}) as well as the following open specification (link: \hyperlink{https://zenodo.org/records/10688871}{https://zenodo.org/records/10688871}). Below in Table \ref{table: comparison}, we present our measured findings for the different prototypes.

\begin{table}[!h]
\centering
\small
\renewcommand{\arraystretch}{1.3}

\begin{tabular*}{\textwidth}{@{\extracolsep{\fill}} l c c c c c @{}}
\toprule
\textbf{Parameter} & 
\textbf{Move} & 
\multicolumn{2}{c}{\textbf{Best Unit Results}} & 
\multicolumn{2}{c}{\textbf{Worst Unit Results}} \\
\cmidrule(lr){3-4} \cmidrule(lr){5-6}
& \textbf{Distance} & \textbf{MPS} & \textbf{Orbray} & \textbf{MPS} & \textbf{Orbray} \\
\midrule\midrule

Repeatability $\alpha$ ($\mu$m RMS) & typ. 30$^\circ$ & 0.65 & 1.58 & 2.12  & 34.52 \\
Repeatability $\beta$ ($\mu$m RMS) & typ. 30$^\circ$ & 0.59  & 1.03  & 0.99  & 44.10  \\
Repeatability (Comb.) ($\mu$m RMS) & typ. 30$^\circ$ & 0.88 & 1.89  & 2.34  & 56.00 \\

\midrule

Small Move Error $\alpha$ ($\mu$m RMS)  & typ. 1$^\circ$ & 2.35 & 2.47 & 3.70 & 13.19 \\
Small Move Error $\beta$ ($\mu$m RMS)   & typ. 1$^\circ$ & 1.29 & 2.09 & 3.99 & 13.99 \\
Small Move Error (Comb.) ($\mu$m RMS)   & typ. 1$^\circ$ & 2.68 & 3.24 & 5.44 & 19.23 \\

\midrule

Backlash $\alpha$      & -- & 1.16$^\circ$ & 0.07$^\circ$ & 7.94$^\circ$ & 10.79$^\circ$ \\
Backlash $\beta$       & -- & 3.69$^\circ$ & 7.66$^\circ$ & 7.42$^\circ$ & 16.67$^\circ$ \\
Angular Tilt           & -- & 0.22$^\circ$ & 0.18$^\circ$ & 0.60$^\circ$ & 0.73$^\circ$ \\

\bottomrule
\end{tabular*}

\caption{Comparison matrix showing the results of the prototypes provided by both manufacturers, MPS and Orbray. Results are shown for the best and worst performing robot units in each set of six units. Combined values for Repeatability and Small Move Error are calculated as the quadrature sum of the $\alpha$ and $\beta$ components.}
\label{table: comparison}
\end{table}

Table \ref{table: comparison} summarizes representative test results, both best- and worst-case values, obtained for the MPS and Orbray prototypes. Overall, the initial performance demonstrated by both systems is highly encouraging and underscores the strong potential of a modular positioner architecture despite the significant mechanical and control challenges it entails.

The more complex mechanics of the Orbray prototype lead to some inconsistencies in the measured XY positioning repeatability, visible in the spread between the best and worst cases listed in the table. The origin of these discrepancies requires further investigation to achieve improved and more stable performance. It is worth noting that the EPFL Orbray prototype has undergone extensive testing and numerous operational cycles, which has contributed to its improved repeatability. In contrast, the LBNL Orbray prototype is relatively new and will benefit from additional burn-in runs to reach comparable stability.

Another interesting parameter that was evaluated for all the tested prototypes is the arc residual or the small move error. For this testing, the robotic arms were commanded to make a small step of 1 degree, and the error of this movement was estimated as the target minus the measured angle. The values for the best units of both prototypes perform relatively well. However, the worst performing units of the Orbray prototype exhibit rather larger values and their performance needs to be improved.

Regarding backlash, simulations of anti-collision path planning for 1000 random tiles of 63 targets each show that accessible area is not reduced when accommodating backlash values of at least 18 degrees or more. Increasing backlash does, however, restrict the minimum allowable proximity between targets, since the non-linear response region of each robot must be expanded and avoided. In this context, the measured backlash values from the tested prototypes are very promising, as they remain well below the critical 18 degree-threshold suggested by the simulations.

The angular tilt measurements show very promising results given that these are the very initial prototypes, and the performances will be further improved on later iterations. Tilt performances are crucial to make sure that the light is properly coupled into the fibers to minimize FRD and throughput losses. Thus, having very small tilt angles is of utmost importance.

Taken together, these results indicate that even at this early stage of development, the modular approach delivers strong performance. Further iterations of the prototypes are expected to refine these metrics substantially.

Future work will include comprehensive testing campaigns covering thermal stability, lifetime performance, and optical throughput and quality for all prototypes.

\begin{acknowledgments}
The authors would like to acknowledge Innosuisse (Funding No.: 101.014 IP-ENG) for supporting this work. Additionally, the authors would like to thank  Dr. Luzius Kronig for the fruitful discussions and his tremendous help. Finally, the authors are very thankful to EPFL workshops for their expertise in making precision parts for the test-benches.

J.H.S., N.W. and D.J.S. are supported by the U.S. Department of Energy (DOE), Office of Science, Office of High-Energy Physics, under Contract No.  DE–AC02–05CH11231.

\end{acknowledgments}

\begin{contribution}

M.G., M.R., J.W., O.P.S., and A.B. conducted the testing and the analysis of the test results. M.R., J.H.S., N.W., S.C., and C.R. developed the mechanics of the prototypes. R.A. and S.P. developed the electronics of the prototypes. D.K. and W.V.S. developed the Trillium design for the Orbray prototypes. Y.K., E.F., S.S, and A.S. developed the Orbray prototypes. J.P.K. and D.S. supervised the work. M.G., M.R., A.B. and J.H.S. wrote the manuscript, and all authors supported with reviewing the manuscript.

\end{contribution}

\bibliography{references, references_1, more_references}{}
\bibliographystyle{aasjournalv7}

\end{document}